\documentclass[prd,showpacs,showkeys,floatfix,onecolumn,amsmath,amssymb,floatfix]{revtex4}
\usepackage{graphicx,color,dcolumn,booktabs,bm}
\usepackage{subfigure}
\usepackage{longtable,lscape}
\usepackage{amssymb}
\usepackage{indentfirst}
\usepackage{epsfig}
\usepackage{feynmf}   
\usepackage{epstopdf}   
\usepackage{slashed}  
\usepackage{cases}
\usepackage{color}
\usepackage{multirow}
\usepackage{graphicx,color,dcolumn,booktabs,bm}
\usepackage[colorlinks,citecolor=blue,anchorcolor=red,menucolor=red, linkcolor=red,filecolor=red,runcolor=red,urlcolor=blue,frenchlinks=red]{hyperref}
\usepackage[marginal]{footmisc}

\begin{document}
\title{Identifying the $\Xi_{b}(6227)$ and $\Sigma_{b}(6097)$ as $P$-wave bottom baryons of $J^P = 3/2^-$}
%

\author{Er-Liang Cui$^1$}
\author{Hui-Min Yang$^1$}
\author{Hua-Xing Chen$^1$}
\email{hxchen@buaa.edu.cn}
\author{Atsushi Hosaka$^{2,3}$}
\email{hosaka@rcnp.osaka-u.ac.jp}

\affiliation{
$^1$School of Physics, Beihang University, Beijing 100191, China \\
$^2$Research Center for Nuclear Physics (RCNP), Osaka University, Ibaraki 567-0047, Japan \\
$^3$Advanced Science Research Center, Japan Atomic Energy Agency (JAEA), Tokai 319-1195, Japan
}

\begin{abstract}
We use the method of QCD sum rules within the framework of heavy quark effective theory to study the mass spectrum of the $\Sigma_{b}(6097)^{\pm}$ and $\Xi_{b}(6227)^{-}$, and use the method of light-cone sum rules still within the heavy quark effective theory to study their decay properties. Our results suggest that they can be well interpreted as $P$-wave bottom baryons with the spin-parity $J^P = 3/2^-$. They belong to the baryon doublet $[\mathbf{6}_F, 2, 1, \lambda]$, where the total and spin angular momenta of the light degree of freedom are $j_l = 2$ and $s_l = 1$, and the orbital angular momentum is between the bottom quark and the two-light-quark system ($\lambda$-type). This doublet contains six bottom baryons, and we predict masses (mass differences) and decay widths of the other four states to be $M_{\Omega_b(3/2^-)} = 6.46 \pm 0.12 {~\rm GeV}$, $\Gamma_{\Omega_b(3/2^-)} = 58{^{+65}_{-33}} {~\rm MeV}$, $M_{\Sigma_b(5/2^-)}-M_{\Sigma_b(3/2^-)}= 13 \pm 5 {~\rm MeV}$, $M_{\Xi_b^{\prime}(5/2^-)}-M_{\Xi_b^{\prime}(3/2^-)} = 12 \pm 5 {~\rm MeV}$, and $M_{\Omega_b(5/2^-)}-M_{\Omega_b(3/2^-)} = 11 \pm 5 {~\rm MeV}$.
We propose to search for them in further LHCb experiments.
\end{abstract}
\maketitle
\pagenumbering{arabic}
%
%
%
\section{Introduction}\label{sec:intro}
%

Recently, the LHCb Collaboration reported their discoveries of two new excited bottom baryons, the $\Xi_{b}(6227)^{-}$ and $\Sigma_{b}(6097)^{\pm}$~\cite{Aaij:2018yqz,Aaij:2018tnn}.
The $\Xi_{b}(6227)^{-}$ was observed in both the $\Lambda^0_b K^-$ and $\Xi^0_b \pi^-$ invariant mass spectra, and the $\Sigma_{b}(6097)^{\pm}$ was observed in the $\Lambda^0_b \pi^\pm$ invariant mass spectrum.
Their masses and decay widths were determined by LHCb to be~\cite{Aaij:2018yqz,Aaij:2018tnn}
\begin{eqnarray}
\nonumber              \Xi_{b}(6227)^{-}:M&=&6226.9 \pm 2.0 \pm 0.3 \pm 0.2 \mbox{ MeV} \, ,
\\ \nonumber                        \Gamma&=&18.1 \pm 5.4 \pm 1.8 \mbox{ MeV} \, ,
\\ \nonumber        \Sigma_{b}(6097)^{+}:M&=&6095.8 \pm 1.7 \pm 0.4 \mbox{ MeV} \, ,
\\ \nonumber                        \Gamma&=&31 \pm 5.5 \pm 0.7 \mbox{ MeV} \, ,
\\ \nonumber        \Sigma_{b}(6097)^{-}:M&=&6098.0 \pm 1.7 \pm 0.5 \mbox{ MeV} \, ,
\\ \                                \Gamma&=&28.9 \pm 4.2 \pm 0.9 \mbox{ MeV} \, .
\end{eqnarray}
The LHCb experiment also measured the following branching ratio~\cite{Aaij:2018yqz}
\begin{eqnarray}
{{\mathcal B}(\Xi_{b}(6227)^{-} \rightarrow \Lambda_b^0 K^-) \over {\mathcal B}(\Xi_{b}(6227)^{-} \rightarrow \Xi_b^0 \pi^-)} \simeq 1 \, .
\label{eq:br}
\end{eqnarray}

The $\Sigma_{b}(6097)^{\pm}$ and $\Xi_{b}(6227)^{-}$ are good candidates of $P$-wave bottom baryons. Besides them, there have been lots of excited heavy baryons observed in various experiments in the past years~\cite{pdg,Yelton:2016fqw,Kato:2016hca,Aaij:2017vbw,Aaij:2017nav}. Their observations quickly attracted lots of interests from theorists~\cite{Chen:2016spr,Cheng:2015iom,Crede:2013sze,Klempt:2009pi,Bianco:2003vb,Korner:1994nh}, but there are only a few Lattice QCD studies~\cite{Padmanath:2013bla,Padmanath:2017lng}. In Refs.~\cite{Chen:2018orb,Chen:2018vuc}, the authors carried out a phenomenological analysis of the $2S$ and $1P$ bottom baryons, and their results suggest that the $\Sigma_{b}(6097)$ and $\Xi_{b}(6227)$ are good candidates of $P$-wave bottom baryons with $J^P = 3/2^-$ or $5/2^-$. This is supported by Ref.~\cite{Yang:2018lzg} by using the quark pair creation ($^3P_0$) model.
In Ref.~\cite{Wang:2018fjm} the authors applied the chiral quark model to study strong decays of $P$-wave bottom baryons, and their results suggest that the $\Sigma_{b}(6097)$ and $\Xi_{b}(6227)$ can be assigned as the $P$-wave bottom baryons belonging to $\mathbf{6}_F$ and having the spin-parity $J^P = 3/2^-$ or $5/2^-$. Non-leptonic decays of $P$-wave bottom baryons were studied in Ref.~\cite{Chua:2018lfa}, and their mass scaling was investigated in Ref.~\cite{Karliner:2018bms}. Besides the $P$-wave bottom baryon interpretation, there also exists the molecular interpretation for the $\Xi_{b}(6227)$~\cite{Huang:2018bed,Yu:2018yxl}.

We have systematically investigated the mass spectrum of excited heavy baryons in Refs.~\cite{Chen:2015kpa,Mao:2015gya,Chen:2016phw,Mao:2017wbz} using the method of QCD sum rules~\cite{Shifman:1978bx,Reinders:1984sr} within the framework of heavy quark effective theory (HQET)~\cite{Grinstein:1990mj,Eichten:1989zv,Falk:1990yz}.
More discussions on heavy mesons and baryons containing a single heavy quark can be found in Refs.~\cite{Bagan:1991sg,Neubert:1991sp,Broadhurst:1991fc,Ball:1993xv,Huang:1994zj,Dai:1996yw,Colangelo:1998ga,Groote:1996em,Zhu:2000py,Lee:2000tb,Huang:2000tn,
Wang:2003zp,Duraes:2007te,Zhou:2014ytp,Zhou:2015ywa}. Especially, in the abstract of Ref.~\cite{Mao:2015gya} we wrote that:``We also study the $SU(3)$ $\mathbf{6}_F$ multiplets by using the baryon multiplets $[\mathbf{6}_F, 0, 1, \lambda]$, $[\mathbf{6}_F, 1, 0, \rho]$ and $[\mathbf{6}_F, 2, 1, \lambda]$, and our results suggest that the $P$-wave bottom baryons $\Sigma_b$, $\Xi^\prime_b$ and $\Omega_b$ have (averaged) masses about 6.0 GeV, 6.2 GeV and 6.4 GeV, respectively.'' Accordingly, our previous results~\cite{Mao:2015gya} suggest that the $\Sigma_{b}(6097)$ and $\Xi_{b}(6227)$ can be well interpreted as $P$-wave bottom baryons belonging to the baryon multiplets $[\mathbf{6}_F, 0, 1, \lambda]$, $[\mathbf{6}_F, 1, 0, \rho]$ and $[\mathbf{6}_F, 2, 1, \lambda]$.

In this paper we shall update our previous QCD sum rule analyses~\cite{Chen:2015kpa,Mao:2015gya}. We shall further study decay properties of $P$-wave bottom baryons to check which baryon multiplet is preferred. To do this we shall use the method of light-cone sum rules within HQET. Similar light-cone sum rule studies were performed in Refs.~\cite{Aliev:2018lcs,Aliev:2018vye,Azizi:2015ksa} to interpret the $\Sigma_{b}(6097)$ and $\Xi_{b}(6227)$ as $P$-wave bottom baryons with $J^P = 3/2^-$, but those studies are in full QCD and not in HQET. In Ref.~\cite{Chen:2017sci} we have systematically investigated the $S$-wave decay properties of $P$-wave charmed baryons, and in this paper we just need to replace the $charm$ quark by the $bottom$ one. We shall also investigate their $D$-wave decay properties. We shall find that the baryon doublet $[\mathbf{6}_F, 2, 1, \lambda]$ can well explain both the masses and decay widths of the $\Sigma_{b}(6097)$ and $\Xi_{b}(6227)$~\cite{Aaij:2018yqz,Aaij:2018tnn}.

This paper is organized as follows. In Sec.~\ref{sec:mass}, we study the mass spectrum of $P$-wave bottom baryons using the method of QCD sum rules within HQET.
In Sec.~\ref{sec:sdecay} and Sec.~\ref{sec:ddecay} we further study their $S$- and $D$-wave decay properties using the method of light-cone sum rules still within HQET.
A short summary is given in Sec.~\ref{sec:summary}.

%
\section{Mass spectrum of $P$-wave bottom baryons}\label{sec:mass}
%

The mass spectrum of $P$-wave heavy baryons have been systematically investigated in Refs.~\cite{Chen:2015kpa,Mao:2015gya} using the method of QCD sum rules within HQET. In this section we update these calculations, and at the same time evaluate the parameters which are needed when calculating their decay widths.

\begin{figure*}[hbt]
\begin{center}
\scalebox{0.7}{\includegraphics{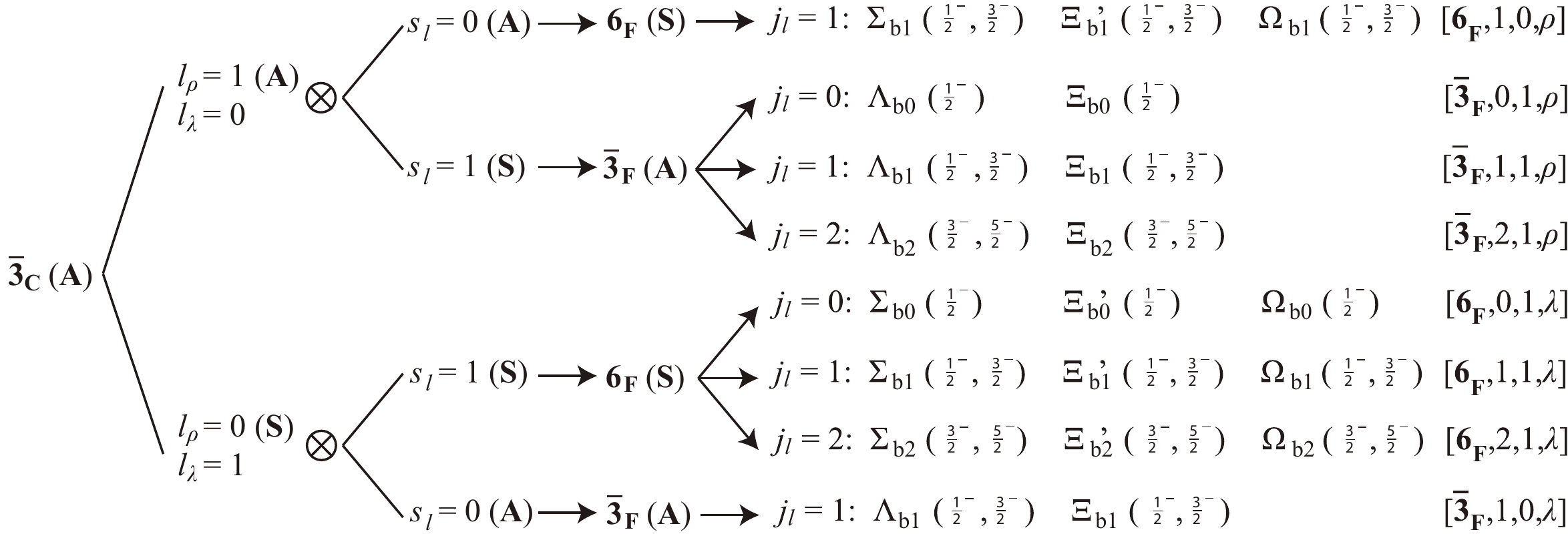}}
\end{center}
\caption{Categorization of $P$-wave bottom baryons.
\label{fig:pwave}}
\end{figure*}

Firstly, let us briefly explain our notations. A bottom baryon contains one bottom quark and two light quarks, and the symmetries between the two light quarks are ($\mathbf{A}=$ antisymmetric and $\mathbf{S}=$ symmetric):
\begin{itemize}

\item The two light quarks have the antisymmetric color structure $\mathbf{\bar 3}_C$ ($\mathbf{A}$).

\item The $P$-wave bottom baryon has one orbital excitation $L = 1$, which can be either between the two light quarks ($\rho$-type, $l_\rho = 1$, $l_\lambda = 0$, $\mathbf{A}$) or between the bottom quark and the two-light-quark system ($\lambda$-type, $l_\rho = 0$, $l_\lambda = 1$, $\mathbf{S}$).

\item The spin of the two light quarks can be either $s_l = 0$ ($\mathbf{A}$) or $s_l = 1$ ($\mathbf{S}$).

\item The $SU(3)$ flavor representation of the two light quarks can be either $\mathbf{\bar 3}_F$ ($\mathbf{A}$) or $\mathbf{6}_F$ ($\mathbf{S}$).

\item The total symmetry of the two light quarks should be antisymmetric due to the Pauli principle.

\end{itemize}
Accordingly, all the $P$-wave bottom baryons can be categorized into eight baryon multiplets, denoted as $[F({\rm flavor}), j_l, s_l, \rho/\lambda]$, where $j_l$ is the total angular momentum of the light components, {\it i.e.}, $j_l = l_\lambda \otimes l_\rho \otimes s_l$. Every multiplet contains several bottom baryons with the total angular momentum $j = j_l \otimes s_Q = | j_l \pm 1/2 |$, where $s_Q = 1/2$ is the spin of the bottom quark. We show the above categorization in Fig.~\ref{fig:pwave}.

All the $P$-wave charmed baryon interpolating fields have been systematically constructed in Ref.~\cite{Chen:2015kpa}, and the bottom baryon fields can be easily obtained by replacing the $charm$ quark field by the $bottom$ one. We use $J^{\alpha_1\cdots\alpha_{j-1/2}}_{j, P, F, j_l, s_l, \rho/\lambda}(x)$ to denote one of these fields, which couples to the bottom baryon $\mathcal{B}$ through
\begin{eqnarray}
\langle 0| J^{\alpha_1\cdots\alpha_{j-1/2}}_{j, P, F, j_l, s_l, \rho/\lambda}(x) | \mathcal{B} \rangle = f_{F, j_l, s_l, \rho/\lambda} u^{\alpha_1\cdots\alpha_{j-1/2}}(x) \, .
\end{eqnarray}
Here $f_{F, j_l, s_l, \rho/\lambda}$ is the decay constant, $P$ is the parity of the bottom baryon, and $u^{\alpha_1\cdots\alpha_j}$ is the relevant spinor. The two-point correlation function at the hadron level can be written as
\begin{eqnarray}
\Pi^{\alpha_1\cdots\alpha_{j-1/2},\beta_1\cdots\beta_{j-1/2}}_{j, P, F, j_l, s_l, \rho/\lambda} (\omega)
&=& i \int d^4 x e^{i k x} \langle 0 | T[J^{\alpha_1\cdots\alpha_{j-1/2}}_{j, P, F, j_l, s_l, \rho/\lambda}(x) \bar J^{\beta_1\cdots\beta_{j-1/2}}_{j, P, F, j_l, s_l, \rho/\lambda}(0)] | 0 \rangle
\label{eq:pole}
\\ \nonumber &=& \mathbb{S} [ g_t^{\alpha_1 \beta_1} \cdots g_t^{\alpha_{j-1/2} \beta_{j-1/2}} ] \times {1 + v\!\!\!\slash \over 2} \times \Pi_{j, P, F, j_l, s_l, \rho/\lambda} (\omega) \, ,
\\ \nonumber &=& \mathbb{S} [ g_t^{\alpha_1 \beta_1} \cdots g_t^{\alpha_{j-1/2} \beta_{j-1/2}} ] \times {1 + v\!\!\!\slash \over 2} \times \Big( {f_{F, j_l, s_l, \rho/\lambda}^{2} \over \overline{\Lambda}_{F, j_l, s_l, \rho/\lambda} - \omega} + \mbox{higher states} \Big) \, .
\end{eqnarray}
Here $\omega$ is the external off-shell energy $\omega = v \cdot k$;
$\mathbb{S} [\cdots]$ denotes symmetrization and subtracting the trace terms in the sets $(\alpha_1 \cdots \alpha_{j-1/2})$ and $(\beta_1 \cdots \beta_{j-1/2})$;
$\overline{\Lambda}_{F,j_l,s_l,\rho/\lambda} \equiv \overline{\Lambda}_{|j_l-1/2|,P,F,j_l,s_l,\rho/\lambda} = \overline{\Lambda}_{j_l+1/2,P,F,j_l,s_l,\rho/\lambda}$ is the sum rule result evaluated at the leading order
\begin{eqnarray}
\overline{\Lambda}_{F, j_l, s_l, \rho/\lambda} \equiv \lim_{m_Q \rightarrow \infty} (m_{{j, P, F, j_l, s_l, \rho/\lambda}} - m_b) \, ,
\end{eqnarray}
where $m_b$ is the {\it bottom} quark mass. We also need to consider the sum rule result at the ${\mathcal O}(1/m_b)$ order so that the mass of the bottom baryon $\mathcal{B}$ can be written as:
\begin{eqnarray}
m_{j,P,F,j_l,s_l,\rho/\lambda} = m_b + \overline{\Lambda}_{F,j_l,s_l,\rho/\lambda} + \delta m_{j,P,F,j_l,s_l,\rho/\lambda} \, .
\label{eq:mass}
\end{eqnarray}
It depends on two free parameters, the threshold value $\omega_c$ and the Borel mass $T$. We refer interested readers to Refs.~\cite{Liu:2007fg,Chen:2015kpa,Mao:2015gya,Chen:2017sci} for detailed explanations of the above equations. Note that the notations used in this paper are the same as those used in Ref.~\cite{Chen:2017sci}, but a bit different from those used in Refs.~\cite{Chen:2015kpa,Mao:2015gya}.

In the present study we update the QCD sum rule results of Ref.~\cite{Mao:2015gya}, and reinvestigate the baryon multiplets $[\mathbf{6}_F, 0, 1, \lambda]$, $[\mathbf{6}_F, 1, 0, \rho]$ and $[\mathbf{6}_F, 2, 1, \lambda]$. As we have found in Ref.~\cite{Mao:2015gya}, there are four baryon multiplets of the flavor $\mathbf{6}_F$ representation, but the other one $[\mathbf{6}_F, 1, 1, \lambda]$ do not give useful QCD sum rule results, so we shall not discuss it in the present study.

From Eq.~(\ref{eq:mass}) we clearly see that the baryon mass depends significantly on the {\it bottom} quark mass, for which we use the $1S$ mass $m_b = 4.66 ^{+0.04}_{-0.03}$ GeV~\cite{pdg2} in the present study. We note that the pole mass $m_b = 4.78 \pm 0.06$ GeV and the $\overline{\rm MS}$ mass $m_b = 4.18 ^{+0.04}_{-0.03}$ GeV~\cite{pdg} are used in some other QCD sum rule studies. This suggests that there are large theoretical uncertainties in our results for the absolute values of the heavy baryon masses, while the mass differences within the same doublet are produced quite well with much less theoretical uncertainties, because they do not depend much on the $bottom$ quark mass~\cite{Zhou:2014ytp,Zhou:2015ywa,Chen:2015kpa,Mao:2015gya}.

Moreover, the baryon mass also moderately depends on the threshold value $\omega_c$, whose uncertainty has been included in our calculation. In the present study we slightly modify the threshold value $\omega_c$ in order to better describe the masses of the $\Sigma_{b}(6097)^\pm$ and $\Xi_{b}(6227)^{-}$ measured by LHCb~\cite{Aaij:2018yqz,Aaij:2018tnn}, so that their decay widths can be better evaluated. We have also finetuned $\omega_c$ in the same multiplet to satisfy the relation $\omega_c(\Omega_c) - \omega_c(\Xi_c^\prime) = \omega_c(\Xi_c^\prime) - \omega_c(\Sigma_c) = 0.15$~GeV, so that the masses of their $\Omega_c$ partner states can be extracted. The obtained results are listed in Table~\ref{tab:pwave}.

\begin{table}[hbt]
\begin{center}
\renewcommand{\arraystretch}{1.5}
\caption{Parameters of the $P$-wave bottom baryons belonging to the baryon multiplets $[\mathbf{6}_F, 0, 1, \lambda]$, $[\mathbf{6}_F, 1, 0, \rho]$ and $[\mathbf{6}_F, 2, 1, \lambda]$. We have slightly modified the threshold value $\omega_c$ compared to those from Ref.~\cite{Mao:2015gya} to better describe the masses of the $\Sigma_{b}(6097)^\pm$ and $\Xi_{b}(6227)^{-}$ measured by LHCb~\cite{Aaij:2018yqz,Aaij:2018tnn}, and finetuned $\omega_c$ in the same multiplet to satisfy the relation $\omega_c(\Omega_c) - \omega_c(\Xi_c^\prime) = \omega_c(\Xi_c^\prime) - \omega_c(\Sigma_c) = 0.15$~GeV. In the last column we have explicitly taken into account the isospin factors, which come from definitions of interpolating fields, {\it i.e.} $f_{\Sigma^+_b} = f_{\Sigma^-_b} = \sqrt2 f_{\Sigma^0_b}$ and $f_{\Xi^{\prime0}_b} = f_{\Xi^{\prime-}_b}$.
Note that there are large theoretical uncertainties in our results for the absolute values of the heavy baryon masses (the 7th column, here we only give the systematic uncertainties), while the mass differences within the same doublet (the 8th column) are produced quite well with much less theoretical uncertainties~\cite{Zhou:2014ytp,Zhou:2015ywa,Chen:2015kpa,Mao:2015gya}.
}
\begin{tabular}{c | c | c | c | c | c c | c | c}
\hline\hline
\multirow{2}{*}{Multiplets} & \multirow{2}{*}{~B~} & $\omega_c$ & ~~~Working region~~~ & ~~~~~~~$\overline{\Lambda}$~~~~~~~ & ~~~Baryons~~~ & ~~~Mass~~~ & ~Difference~ & $f$
\\
  &  & (GeV) & (GeV) & (GeV) & ($j^P$) & (GeV) & (MeV) & (GeV$^{4}$)
\\ \hline\hline
\multirow{3}{*}{$[\mathbf{6}_F, 0, 1, \lambda]$} & $\Sigma_b$ & 1.75 & $0.30< T < 0.33$ & $1.29 \pm 0.08$ & $\Sigma_b(1/2^-)$ & $6.09 \pm 0.10$ & -- & $0.085 \pm 0.017~(\Sigma^-_b(1/2^-))$
\\ \cline{2-9}
& $\Xi^\prime_b$ & 1.90 & $0.30< T < 0.34$ & $1.44 \pm 0.08$ & $\Xi^\prime_b(1/2^-)$ & $6.25 \pm 0.10$ & -- & $0.077 \pm 0.016~(\Xi^{\prime-}_b(1/2^-))$
\\ \cline{2-9}
& $\Omega_b$ & 2.05 & $0.29< T < 0.35$ & $1.59 \pm 0.08$ & $\Omega_b(1/2^-)$ & $6.40 \pm 0.11$ & -- & $0.143 \pm 0.030~(\Omega^-_b(1/2^-))$
\\ \hline
\multirow{6}{*}{$[\mathbf{6}_F, 1, 0, \rho]$} & \multirow{2}{*}{$\Sigma_b$} & \multirow{2}{*}{1.87} & \multirow{2}{*}{$0.31< T < 0.34$} & \multirow{2}{*}{$1.35 \pm 0.09$} & $\Sigma_b(1/2^-)$ & $6.10 \pm 0.11$ & \multirow{2}{*}{$3 \pm 1$} & $0.087 \pm 0.018~(\Sigma^-_b(1/2^-))$
\\ \cline{6-7}\cline{9-9}
& & & & & $\Sigma_b(3/2^-)$ & $6.10 \pm 0.10$ & &$0.050 \pm 0.011~(\Sigma^-_b(3/2^-))$
\\ \cline{2-9}
& \multirow{2}{*}{$\Xi^\prime_b$} & \multirow{2}{*}{2.02} & \multirow{2}{*}{$0.29< T < 0.36$} & \multirow{2}{*}{$1.49 \pm 0.09$} & $\Xi^\prime_b(1/2^-)$ & $6.24 \pm 0.11$ & \multirow{2}{*}{$3 \pm 1$} & $0.080 \pm 0.016~(\Xi^{\prime-}_b(1/2^-))$
\\ \cline{6-7}\cline{9-9}
& & & & & $\Xi^\prime_b(3/2^-)$ & $6.24 \pm 0.11$ & &$0.046 \pm 0.009~(\Xi^{\prime-}_b(3/2^-))$
\\ \cline{2-9}
& \multirow{2}{*}{$\Omega_b$} & \multirow{2}{*}{2.17} & \multirow{2}{*}{$0.33< T < 0.38$} & \multirow{2}{*}{$1.67 \pm 0.09$} & $\Omega_b(1/2^-)$ & $6.42 \pm 0.11$ & \multirow{2}{*}{$3 \pm 1$} & $0.155 \pm 0.030~(\Omega^-_b(1/2^-))$
\\ \cline{6-7}\cline{9-9}
& & & & & $\Omega_b(3/2^-)$ & $6.42 \pm 0.11$ & &$0.090 \pm 0.017~(\Omega^-_b(3/2^-))$
\\ \hline
\multirow{6}{*}{$[\mathbf{6}_F, 2, 1, \lambda]$} & \multirow{2}{*}{$\Sigma_b$} & \multirow{2}{*}{1.84} & \multirow{2}{*}{$0.30< T < 0.34$} & \multirow{2}{*}{$1.29 \pm 0.09$} & $\Sigma_b(3/2^-)$ & $6.10 \pm 0.12$ & \multirow{2}{*}{$13 \pm 5$} & $0.102 \pm 0.022~(\Sigma^-_b(3/2^-))$
\\ \cline{6-7}\cline{9-9}
& & & & & $\Sigma_b(5/2^-)$ & $6.11 \pm 0.12$ & &$0.045 \pm 0.010~(\Sigma^-_b(5/2^-))$
\\ \cline{2-9}
& \multirow{2}{*}{$\Xi^\prime_b$} & \multirow{2}{*}{1.99} & \multirow{2}{*}{$0.30< T < 0.36$} & \multirow{2}{*}{$1.45 \pm 0.09$} & $\Xi^\prime_b(3/2^-)$ & $6.27 \pm 0.12$ & \multirow{2}{*}{$12 \pm 5$} & $0.099 \pm 0.021~(\Xi^{\prime-}_b(3/2^-))$
\\ \cline{6-7}\cline{9-9}
& & & & & $\Xi^\prime_b(5/2^-)$ & $6.29 \pm 0.11$ & &$0.044 \pm 0.009~(\Xi^{\prime-}_b(5/2^-))$
\\ \cline{2-9}
& \multirow{2}{*}{$\Omega_b$} & \multirow{2}{*}{2.14} & \multirow{2}{*}{$0.32< T < 0.38$} & \multirow{2}{*}{$1.62 \pm 0.09$} & $\Omega_b(3/2^-)$ & $6.46 \pm 0.12$ & \multirow{2}{*}{$11 \pm 5$} & $0.194 \pm 0.038~(\Omega^-_b(3/2^-))$
\\ \cline{6-7}\cline{9-9}
& & & & & $\Omega_b(5/2^-)$ & $6.47 \pm 0.12$ & &$0.087 \pm 0.017~(\Omega^-_b(3/2^-))$
\\ \hline \hline
\end{tabular}
\label{tab:pwave}
\end{center}
\end{table}

%
\section{$S$-wave decay properties of $P$-wave bottom baryons}\label{sec:sdecay}
%

In the previous section we studied the mass spectrum of $P$-wave bottom baryons, and found that the masses of the $\Sigma_{b}(6097)^\pm$ and $\Xi_{b}(6227)^{-}$~\cite{Aaij:2018yqz,Aaij:2018tnn} are consistent with those of the $P$-wave bottom baryons belonging to the baryon multiplets $[\mathbf{6}_F, 1, 0, \rho]$, $[\mathbf{6}_F, 0, 1, \lambda]$, and $[\mathbf{6}_F, 2, 1, \lambda]$. However, the uncertainties of our sum rule calculations are not so small, preventing us to differentiate these multiplets, while their decay properties do depend significantly on their internal structures~\cite{Cheng:2006dk,Zhong:2007gp,Nagahiro:2016nsx}.

In this section we further study their decay properties to check which baryon multiplet is preferred. We shall study their $S$-wave decays into ground-state bottom baryons accompanied by a pseudoscalar meson ($\pi$ or $K$), including both the two-body and three-body decays which are kinematically allowed. Their $S$-wave decays into ground-state bottom baryons accompanied by a vector meson ($\rho$ or $K^*$) are kinematically forbidden.

To do this we use the method of light-cone sum rules within HQET. Actually, the decays of $P$-wave charmed baryons have been systematically investigated in Ref.~\cite{Chen:2017sci} using the same approach. In this paper we just need to replace the $charm$ quark by the $bottom$ one, and reinvestigate the following decay channels:
\begin{eqnarray}
&(k)& {\bf \Gamma\Big[} \Sigma_b(1/2^-) \rightarrow \Lambda_b(1/2^+) + \pi {\Big ]}
= {\bf \Gamma\Big[} \Sigma_b^{-}(1/2^-) \rightarrow \Lambda_b^{0}(1/2^+) + \pi^- {\Big ]} \, ,
\\ &(l)& {\bf \Gamma\Big[}\Sigma_b(1/2^-) \rightarrow \Sigma_b(1/2^+) + \pi{\Big ]}
= 2 \times {\bf \Gamma\Big[}\Sigma_b^{-}(1/2^-) \rightarrow \Sigma_b^{0}(1/2^+) + \pi^-{\Big ]} \, ,
\\ &(m)& {\bf \Gamma\Big[}\Xi_b^{\prime}(1/2^-) \rightarrow \Xi_b(1/2^+) + \pi{\Big ]}
= {3\over2} \times {\bf \Gamma\Big[}\Xi_b^{\prime -}(1/2^-) \rightarrow \Xi_b^{0}(1/2^+) + \pi^-{\Big ]} \, ,
\\ &(n)& {\bf \Gamma\Big[} \Xi_b^{\prime}(1/2^-) \rightarrow \Lambda_b(1/2^+) + K {\Big ]}
= {\bf \Gamma\Big[} \Xi_b^{\prime -}(1/2^-) \rightarrow \Lambda_b^{0}(1/2^+) + K^- {\Big ]} \, ,
\\ &(o)& {\bf \Gamma\Big[}\Xi_b^{\prime}(1/2^-) \rightarrow \Xi_b^{\prime}(1/2^+) + \pi{\Big ]}
= {3\over2} \times  {\bf \Gamma\Big[}\Xi_b^{\prime -}(1/2^-) \rightarrow \Xi_b^{\prime0}(1/2^+) + \pi^-{\Big ]} \, ,
\\ &(p)& {\bf \Gamma\Big[}\Xi_b^{\prime}(1/2^-) \rightarrow \Sigma_b(1/2^+) + K{\Big ]}
= 3 \times {\bf \Gamma\Big[}\Xi_b^{\prime -}(1/2^-) \rightarrow \Sigma_b^{0}(1/2^+) + K^-{\Big ]} \, ,
\\ &(q)& {\bf \Gamma\Big[}\Omega_b(1/2^-) \rightarrow \Xi_b(1/2^+) + K{\Big ]}
= 2 \times {\bf \Gamma\Big[}\Omega_b^{-}(1/2^-) \rightarrow \Xi_b^{0}(1/2^+) + K^-{\Big ]} \, ,
\\ &(r)& {\bf \Gamma\Big[}\Omega_b(1/2^-) \rightarrow \Xi_b^{\prime}(1/2^+) + K{\Big ]}
= 2 \times {\bf \Gamma\Big[}\Omega_b^{-}(1/2^-) \rightarrow \Xi_b^{\prime0}(1/2^+) + K^-{\Big ]} \, ,
\\&(s)& {\bf \Gamma\Big[}\Sigma_b(3/2^-) \rightarrow \Sigma_b^{*}(3/2^+) + \pi {\Big ]}
= 2 \times {\bf \Gamma\Big[}\Sigma_b^{-}(3/2^-) \rightarrow \Sigma_b^{*0}(3/2^+) + \pi^- {\Big ]} \, ,
\\ &(t)& {\bf \Gamma\Big[}\Xi_b^{\prime}(3/2^-) \rightarrow \Xi_b^{*}(3/2^+) + \pi {\Big ]}
= {3\over2} \times {\bf \Gamma\Big[}\Xi_b^{\prime-}(3/2^-) \rightarrow \Xi_b^{*0}(3/2^+) + \pi^- {\Big ]} \, ,
\end{eqnarray}
\begin{eqnarray}
&(u)& {\bf \Gamma\Big[}\Xi_b^{\prime}(3/2^-) \rightarrow \Sigma_b^{*}(3/2^+) + K \rightarrow \Lambda_b(1/2^+) + \pi + K {\Big ]}
\\ \nonumber &=& 3 \times {\bf \Gamma\Big[}\Xi_b^{\prime-}(3/2^-) \rightarrow \Sigma_b^{*0}(3/2^+) + K^- \rightarrow \Lambda_b^{0}(3/2^+) + \pi^0 + K^- {\Big ]} \, ,
\\ &(v)& {\bf \Gamma\Big[}\Omega_b(3/2^-) \rightarrow \Xi_b^{*}(3/2^+) + K {\Big ]} = 2 \times {\bf \Gamma\Big[}\Omega_b^{-}(3/2^-) \rightarrow \Xi_b^{*0}(3/2^+) + K^- {\Big ]} \, .
\end{eqnarray}
We can calculate their decay widths through the following Lagrangians
\begin{eqnarray}
\mathcal{L}_{X_b({1/2}^-) \rightarrow Y_b({1/2}^+) P} &=& g {\bar X_b}(1/2^-) Y_b(1/2^+) P \, ,
\\ \mathcal{L}_{X_b({3/2}^-) \rightarrow Y_b({3/2}^+) P} &=& g {\bar X_{b}^{\mu}}(3/2^-) Y_{b}^{\mu}(3/2^+) P \, ,
\end{eqnarray}
where $X_b^{(\mu)}$, $Y_b^{(\mu)}$, and $P$ denotes the excited bottom baryon, ground-state bottom baryon, and pseudoscalar meson, respectively.

As an example, we shall study the $S$-wave decay of $\Sigma_b^-({1/2}^-)$ belonging to $[\mathbf{6}_F, 1, 0, \rho]$ into $\Sigma_b^0(1/2^+)$ and $\pi^-(0^-)$ in the next subsection. Then we shall investigate the three baryon multiplets, $[\mathbf{6}_F, 1, 0, \rho]$, $[\mathbf{6}_F, 0, 1, \lambda]$, and $[\mathbf{6}_F, 2, 1, \lambda]$, separately in the following subsections. The results are summarized in Table~\ref{tab:sdecay}.

\begin{table}[hbt]
\begin{center}
\renewcommand{\arraystretch}{1.5}
\caption{$S$-wave decay properties of the $P$-wave bottom baryons belonging to the baryon multiplets $[\mathbf{6}_F, 0, 1, \lambda]$, $[\mathbf{6}_F, 1, 0, \rho]$ and $[\mathbf{6}_F, 2, 1, \lambda]$.}
\begin{tabular}{c | l | c | c}
\hline\hline
Multiplets & ~~~$S$-wave decay channels~~~ & ~$g$ ~ & $S$-wave decay width (MeV)
\\ \hline\hline
\multirow{6}{*}{$[\mathbf{6}_F,1,0,\rho]$}       & (l) $\Sigma_b({1\over2}^-)\to \Sigma_b({1\over2}^+) \pi$  & $3.41^{+1.74}_{-1.33}$  & $ 850^{+1100}_{-540}$
\\
                                                 & (o) $\Xi_b^{\prime}({1\over2}^-)\to \Xi_b^{\prime}({1\over2}^+) \pi$ & $2.31^{+1.13}_{-0.87}$ & $310^{+370}_{-190}$
\\
                                                 & (s) $\Sigma_b({3\over2}^-)\to \Sigma_b^{*}({3\over2}^+) \pi$ & $2.28^{+1.16}_{-0.89}$ & $ 350^{+440}_{-220}$
\\
                                                 & (t) $\Xi_b^{\prime}({3\over2}^-)\to \Xi_b^{*}({3\over2}^+) \pi$ & $1.54^{+0.75}_{-0.58}$ & $ 130^{+150}_{-80}$
\\
                                                 & (u) $\Xi_b^{\prime}({3\over2}^-)\to \Sigma_b^{*}({3\over2}^+) K \to \Lambda_b({1\over2}^+) \pi K$ & $2.10^{+1.07}_{-0.79}$ & $ 0.029^{+0.036}_{-0.017}$
\\
                                                 & (v) $\Omega_b({3\over2}^-) \to \Xi_b^*({3\over2}^+) K$ & $2.72^{+1.29}_{-0.96}$ & --
\\ \hline
\multirow{4}{*}{$[\mathbf{6}_F,0,1,\lambda]$}    & (k) $\Sigma_b({1\over2}^-)\to \Lambda_b({1\over2}^+) \pi$ & $4.70^{+2.39}_{-1.84}$ & $ 1400^{+1800}_{-900}$
\\
                                                 & (m) $\Xi_b^{\prime}({1\over2}^-)\to \Xi_b({1\over2}^+) \pi$ & $3.40^{+1.69}_{-1.30}$ & $1000^{+1300}_{-630}$
\\
                                                 & (n) $\Xi_b^{\prime}({1\over2}^-)\to \Lambda_b({1\over2}^+) K$ & $4.56^{+2.35}_{-1.74}$ & $1000^{+1300}_{-620}$
\\
                                                 & (q) $\Omega_b({1\over2}^-)\to \Xi_b({1\over2}^+) K$ & $6.38^{+3.16}_{-2.35}$ & $3900^{+4900}_{-2400}$
\\ \hline
\multirow{4}{*}{$[\mathbf{6}_F,2,1,\lambda]$}    & (s) $\Sigma_b({3\over2}^-)\to \Sigma_b^{*}({3\over2}^+) \pi$ & $0.014^{+0.008}_{-0.007}$ &
                                                 $0.013^{+0.019}_{-0.010}$
\\
                                                 & (t) $\Xi_b^{\prime}({3\over2}^-)\to \Xi_b^{*}({3\over2}^+) \pi$ & $0.009^{+0.005}_{-0.005}$ &$ 0.004^{+0.006}_{-0.003}$
\\
                                                 & (u) $\Xi_b^{\prime}({3\over2}^-)\to \Sigma_b^{*}({3\over2}^+) K \to \Lambda_b({1\over2}^+) \pi K$ & $0.006^{+0.010}_{-0.006}$ & $2^{+14}_{-2} \times 10^{-7}$
\\
                                                 & (v) $\Omega_b({3\over2}^-)\to \Xi_b^{*}({3\over2}^+) K$ & $0.007^{+0.012}_{-0.007}$ & $ 0.001^{+0.008}_{-0.001}$
\\ \hline \hline
\end{tabular}
\label{tab:sdecay}
\end{center}
\end{table}

\subsection{$\Sigma_b^-({1/2}^-)$ of $[\mathbf{6}_F, 1, 0, \rho]$ decaying into $\Sigma_b^0(1/2^+)$ and $\pi^-(0^-)$}

As an example, in this subsection we study the $S$-wave decay of the $\Sigma_b^-({1/2}^-)$ belonging to $[\mathbf{6}_F, 1, 0, \rho]$ into $\Sigma_b^0(1/2^+)$ and $\pi^-(0^-)$. To do this we consider the three-point correlation function:
\begin{eqnarray}
\Pi(\omega, \, \omega^\prime) &=& \int d^4 x~e^{-i k \cdot x}~\langle 0 | J_{1/2,-,\Sigma_b^-,1,0,\rho}(0) \bar J_{\Sigma_b^{0}}(x) | \pi^- \rangle
\\ \nonumber &=& {1+v\!\!\!\slash\over2} G_{\Sigma_b^-[{1\over2}^-] \rightarrow \Sigma_b^{0}\pi^-} (\omega, \omega^\prime) \, ,
\end{eqnarray}
where $k^\prime = k + q$, $\omega^\prime = v \cdot k^\prime$, and $\omega = v \cdot k$.

At the hadronic level, $G_{\Sigma_b^-[{1\over2}^-] \rightarrow \Sigma_b^{0}\pi^-}$ has the following pole terms from the double dispersion relation:
\begin{eqnarray}
G_{\Sigma_b^-[{1\over2}^-] \rightarrow \Sigma_b^{0}\pi^-} (\omega, \omega^\prime) &=& g_{\Sigma_b^-[{1\over2}^-] \rightarrow \Sigma_b^{0}\pi^-} \times { f_{\Sigma_b^-[{1\over2}^-]} f_{\Sigma_b^{0}} \over (\bar \Lambda_{\Sigma_b^-[{1\over2}^-]} - \omega^\prime) (\bar \Lambda_{\Sigma_b^{0}} - \omega)} \, , \label{G0C}
\end{eqnarray}
while it can also be calculated at the quark and gluon level using the method of operator product expansion. The sum rules for the charmed baryon decay $\Sigma_c^0({1/2}^-) \rightarrow \Sigma_c^+(1/2^+) \pi^-$ have been calculated in Ref.~\cite{Chen:2017sci}, and the sum rules for the bottom baryon decay $\Sigma_b^-({1/2}^-) \rightarrow \Sigma_b^0(1/2^+) \pi^-$ are almost the same:
\begin{eqnarray}
\label{eq:sumrule}
&& G_{\Sigma_b^-[{1\over2}^-] \rightarrow \Sigma_b^0\pi^-} (\omega, \omega^\prime)
= g_{\Sigma_b^-[{1\over2}^-] \rightarrow \Sigma_b^{0}\pi^-} \times { f_{\Sigma_b^-[{1\over2}^-]} f_{\Sigma_b^{0}} \over (\bar \Lambda_{\Sigma_b^-[{1\over2}^-]} - \omega^\prime) (\bar \Lambda_{\Sigma_b^{0}} - \omega)}
\\ \nonumber &=& \int_0^\infty dt \int_0^1 du e^{i (1-u) \omega^\prime t} e^{i u \omega t} \times 8 \times \Big (
\frac{3 f_\pi m_\pi^2}{4 \pi^2 t^4 (m_u + m_d)} \phi_{3;\pi}^p(u)
+ \frac{i f_\pi m_\pi^2 v \cdot q}{8 \pi^2 t^3 ( m_u + m_d )} \phi_{3;\pi}^\sigma(u)
\\ \nonumber &&
- \frac{i f_\pi}{16 t v \cdot q} \langle \bar q q \rangle \psi_{4;\pi}(u)
- \frac{i f_\pi t}{256 v\cdot q} \langle g_s \bar q \sigma G q\rangle \psi_{4;\pi}(u)
\Big ) \, .
\end{eqnarray}
In the above expression the radiative corrections are not taken into account, which are usually not important in QCD sum rule studies~\cite{Albuquerque:2018jkn}.
The explicit forms of the light-cone distribution amplitudes contained in this equation can be found in Refs.~\cite{Ball:1998je,Ball:2006wn,Ball:2004rg,Ball:1998kk,Ball:1998sk,Ball:1998ff,Ball:2007rt,Ball:2007zt}, where their values can also be found. In this paper we work at the renormalization scale 2 GeV, and use the following values for various condensates~\cite{pdg,Yang:1993bp,Hwang:1994vp,Ovchinnikov:1988gk,Jamin:2002ev,Ioffe:2002be,Narison:2002pw,Gimenez:2005nt,colangelo}:
%
\begin{eqnarray}
\nonumber && \langle \bar qq \rangle = - (0.24 \pm 0.01 \mbox{ GeV})^3 \, ,
\\ \nonumber && \langle \bar ss \rangle = (0.8\pm 0.1)\times \langle\bar qq \rangle \, ,
\\ &&\langle g_s^2GG\rangle =(0.48\pm 0.14) \mbox{ GeV}^4\, ,
\label{eq:condensates}
\\ \nonumber && \langle g_s \bar q \sigma G q \rangle = M_0^2 \times \langle \bar qq \rangle\, ,
\\ \nonumber && \langle g_s \bar s \sigma G s \rangle = M_0^2 \times \langle \bar ss \rangle\, ,
\\ \nonumber && M_0^2= 0.8 \mbox{ GeV}^2\, .
\end{eqnarray}
Then we perform the Wick rotations and double Borel transformation to Eq.~(\ref{eq:sumrule}) with $\omega$ and $\omega^\prime$ replaced by $T_1$ and $T_2$:
\begin{eqnarray}
&& g_{\Sigma_b^-[{1\over2}^-] \rightarrow \Sigma_b^{0}\pi^-} f_{\Sigma_b^-[{1\over2}^-]} f_{\Sigma_b^{0}} e^{- {\bar \Lambda_{\Sigma_b^-[{1\over2}^-]} \over T_1}} e^{ - {\bar \Lambda_{\Sigma_b^{0}} \over T_2}}
\\ \nonumber &=& 8 \times \Big ( \frac{3 i f_\pi m_\pi^2}{4 \pi^2 (m_u + m_d)} T^5 f_4({\omega_c \over T}) \phi_{3;\pi}^p(u_0)
+ \frac{i f_\pi m_\pi^2}{8 \pi^2 (m_u + m_d)} T^5 f_4({\omega_c \over T}) {d \phi_{3;\pi}^\sigma(u_0) \over du}
\\ \nonumber &&
+ \frac{i f_\pi}{16} {\langle \bar q q \rangle} T f_0({\omega_c \over T}) {\int_0^{u_0} \psi_{4;\pi}(u)du}
- \frac{i f_\pi}{256} {\langle g_s \bar q \sigma G q \rangle} {1 \over T} {\int_0^{u_0} \psi_{4;\pi}(u)du}
 \Big )
\, ,
\end{eqnarray}
where $u_0 = {T_1 \over T_1 + T_2}$, $T = {T_1 T_2 \over T_1 + T_2}$ and $f_n(x) = 1 - e^{-x} \sum_{k=0}^n {x^k \over k!}$.

We work at the symmetric point $T_1 = T_2 = 2T$, and choose $\omega_c = 1.60$ GeV to be the average of the threshold values of the $\Sigma_b(1/2^-)$ and $\Sigma_b^{0}$ mass sum rules. The coupling constant $g_{\Sigma_b^-[{1\over2}^-] \rightarrow \Sigma_b^{0}\pi^-}$ and the width of the $\Sigma_b^-({1/2}^-) \rightarrow \Sigma_b^0(1/2^+) \pi^-$ decay are extracted to be
\begin{eqnarray}
g_{\Sigma_b^-[{1\over2}^-] \rightarrow \Sigma_b^{0}\pi^-} &=& 3.41~{^{+1.74}_{-1.33}} = 3.41~{^{+0.03 }_{-0.02}}~{^{+1.06}_{-0.85}}~{^{+1.05}_{-0.76}}~{^{+0.89}_{-0.69}} \, ,
\\
\Gamma_{\Sigma_b^-[{1\over2}^-] \rightarrow \Sigma_b^{0}\pi^-} &=& 850~{^{+1100}_{-540}} {\rm~MeV} \, ,
\end{eqnarray}
where the uncertainties come from the Borel mass, the parameters of the $\Sigma_b^{0}$, the parameters of the $\Sigma_b^-[{1\over2}^-]$, and various quark masses and condensates listed in Eq.~(\ref{eq:condensates}), respectively.
For completeness, we show $g_{\Sigma_b^-[{1\over2}^-] \rightarrow \Sigma_b^{0}\pi^-}$ in Fig.~\ref{fig:610rho}(a) as a function of the Borel mass $T$.
We find that it only slightly depends on the Borel mass. Moreover, its uncertainty due to the Borel mass is quite small, so the narrow Borel windows chosen in the present study are acceptable and do not affect the results significantly.

In the following subsections we shall follow the same procedures to separately study the three baryon multiplets, $[\mathbf{6}_F, 1, 0, \rho]$, $[\mathbf{6}_F, 0, 1, \lambda]$, and $[\mathbf{6}_F, 2, 1, \lambda]$.

\subsection{The baryon doublet $[\mathbf{6}_F, 1, 0, \rho]$}

\begin{figure}[htb]
\begin{center}
\subfigure[]{
\scalebox{0.6}{\includegraphics{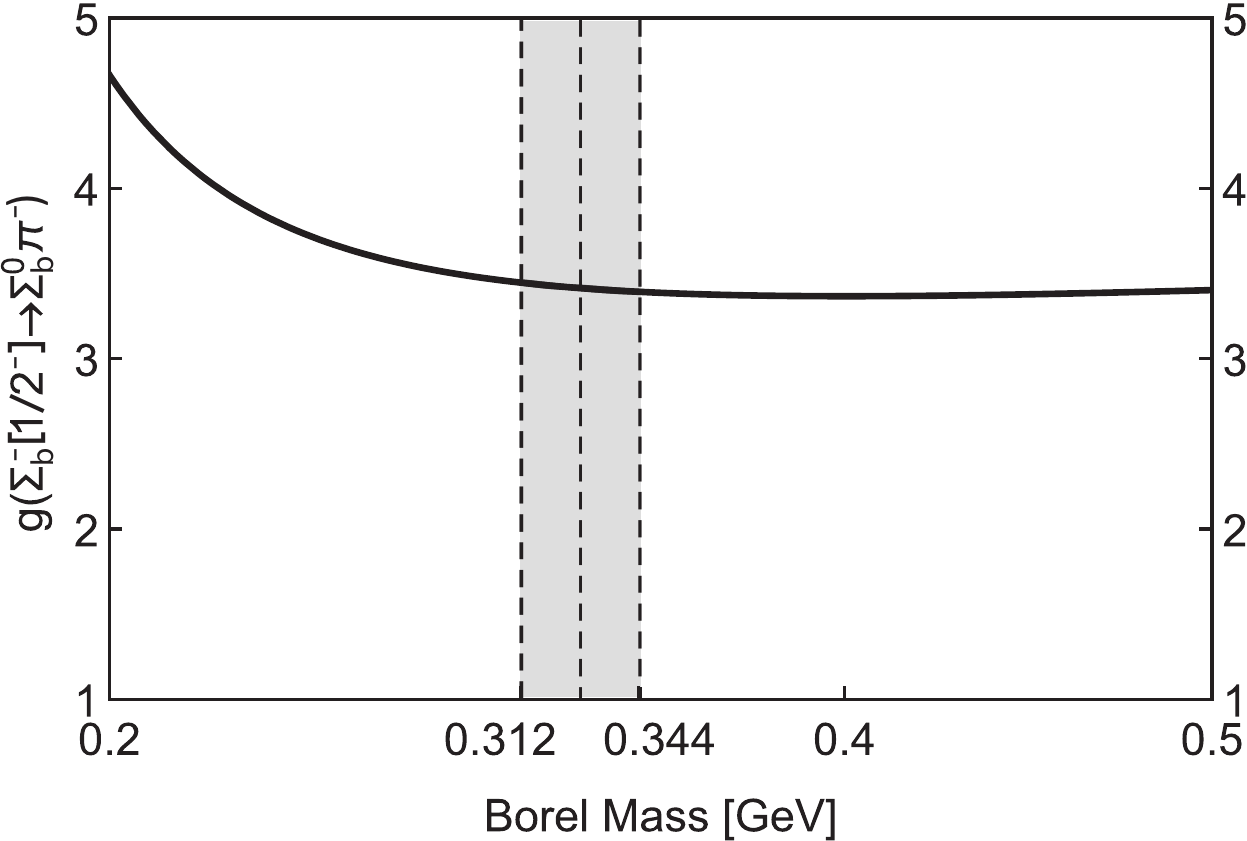}}}
~~~~~
\subfigure[]{
\scalebox{0.6}{\includegraphics{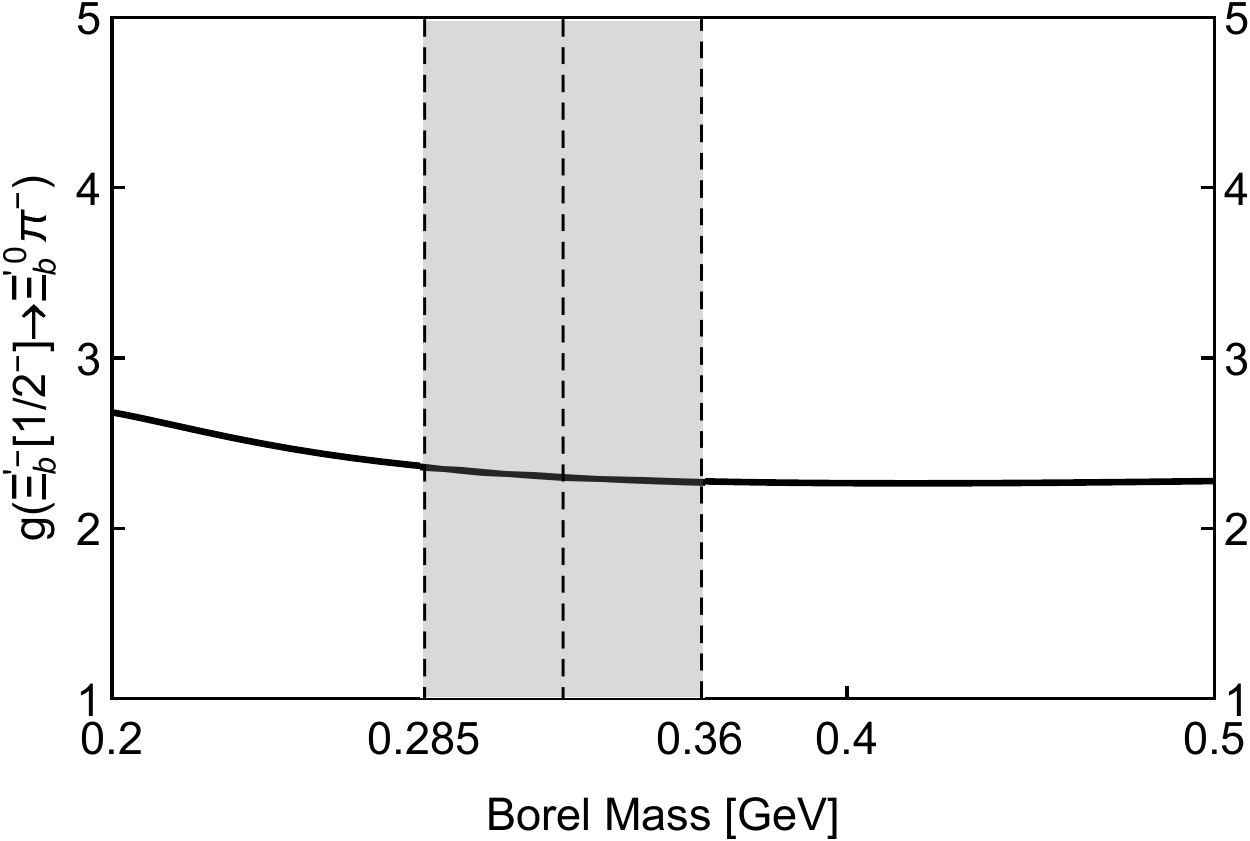}}}
\\
\subfigure[]{
\scalebox{0.6}{\includegraphics{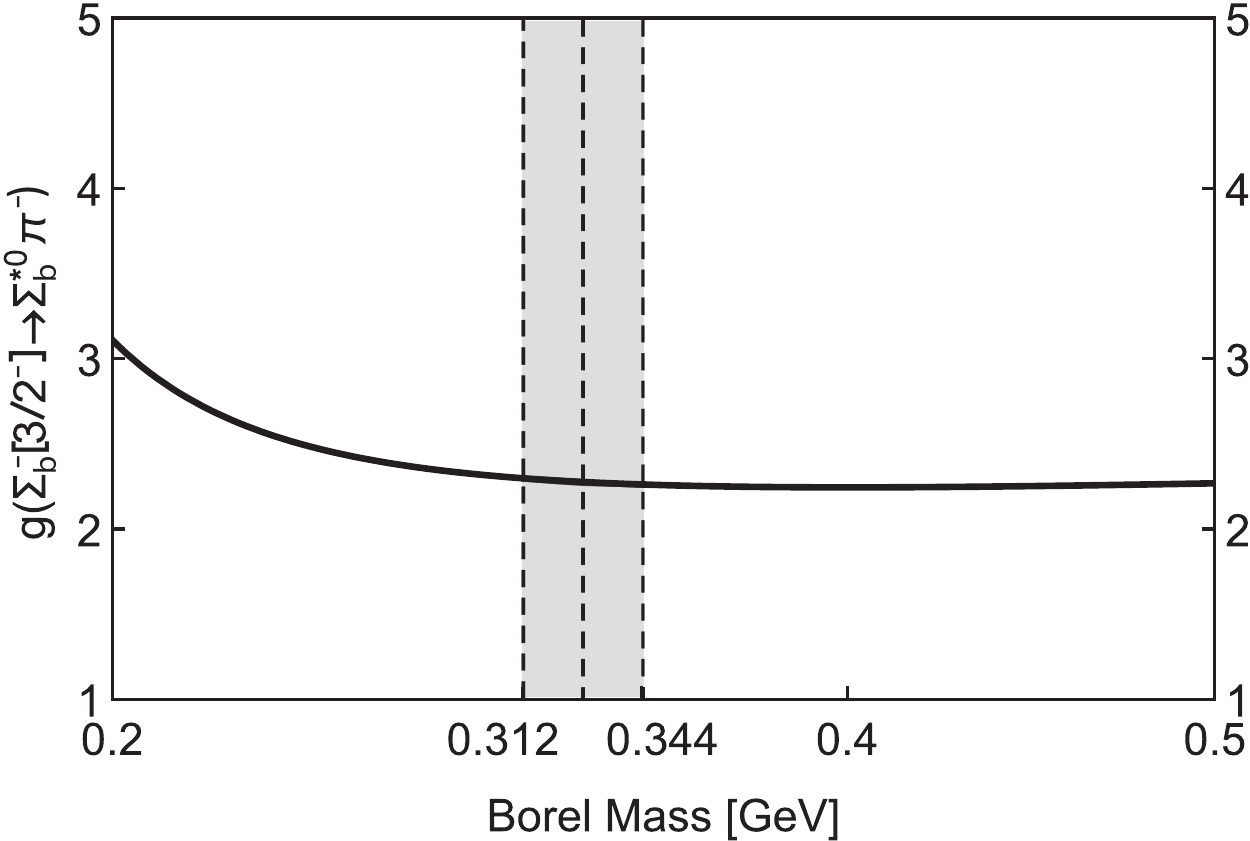}}}
~~~~~
\subfigure[]{
\scalebox{0.6}{\includegraphics{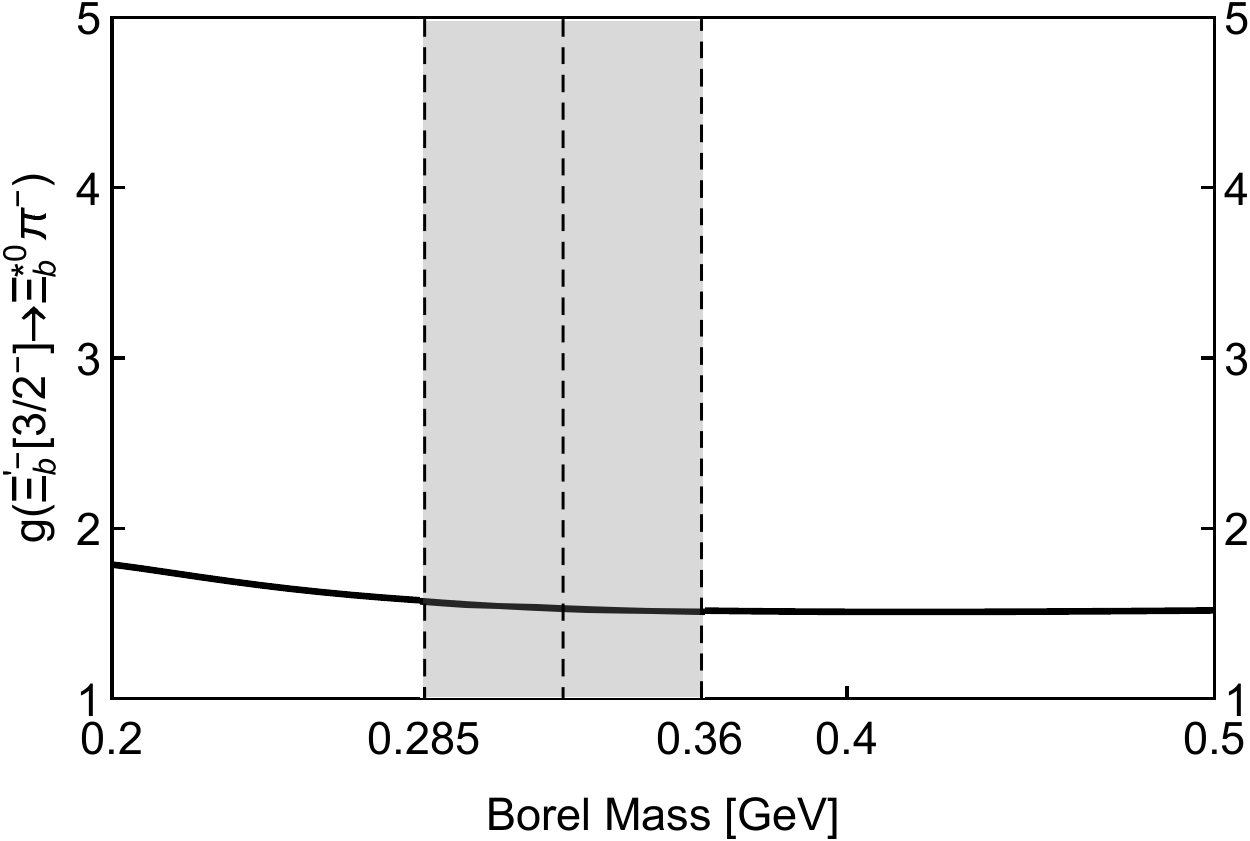}}}
\\
\subfigure[]{
\scalebox{0.6}{\includegraphics{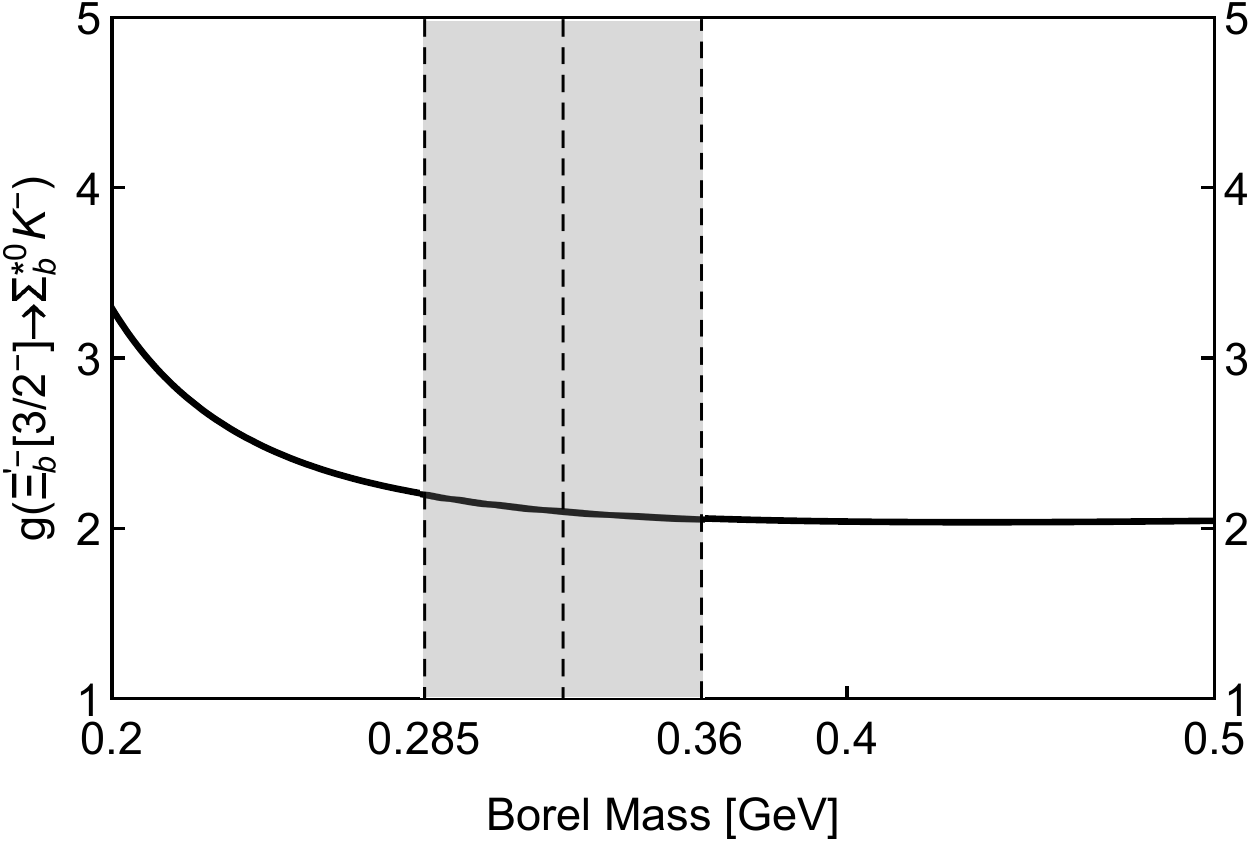}}}
~~~~~
\subfigure[]{
\scalebox{0.6}{\includegraphics{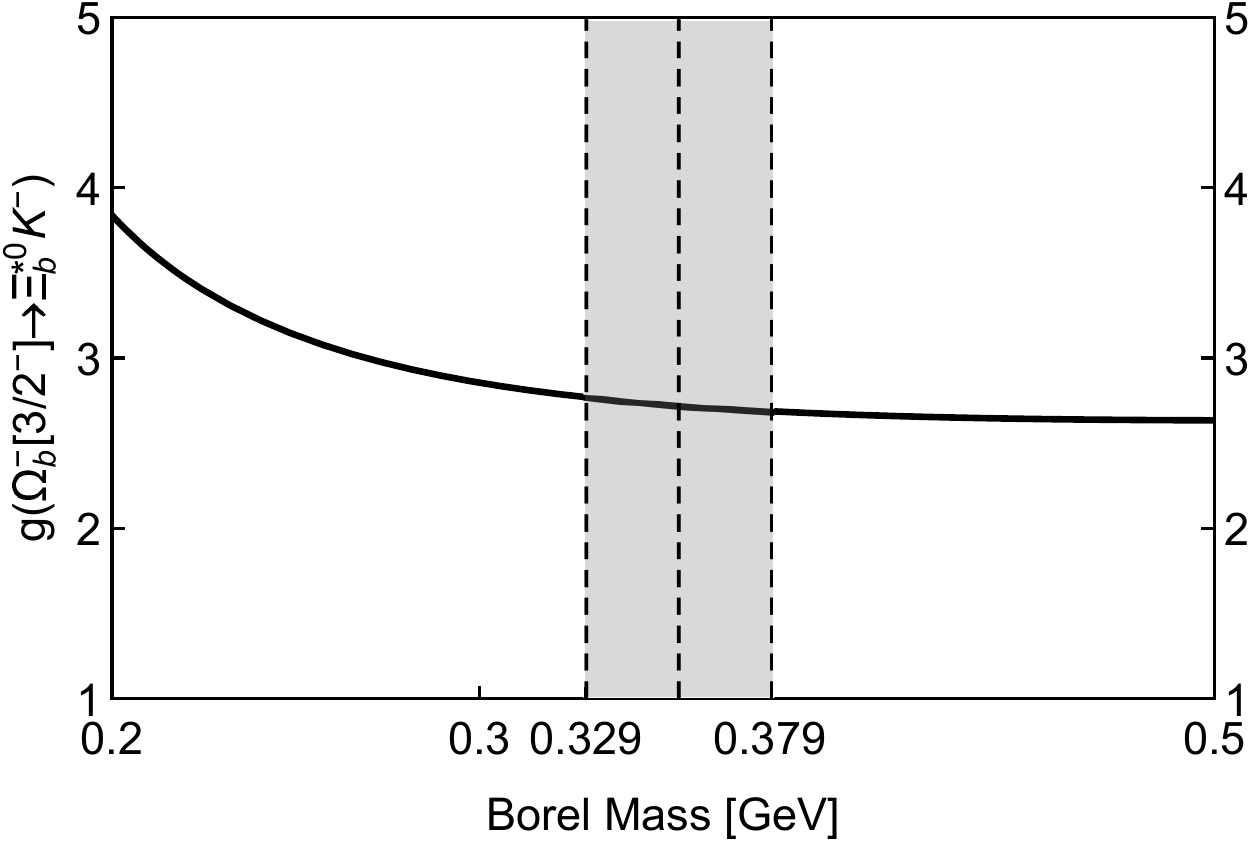}}}
\end{center}
\caption{The $S$-wave decay coupling constants (a) $g_{\Sigma_b^-[{1\over2}^-] \rightarrow \Sigma_b^{0} \pi^-}$, (b) $g_{\Xi_b^{\prime-}[{1\over2}^-] \rightarrow \Xi_b^{\prime0}\pi^-}$, (c) $g_{\Sigma_b^{-}[{3\over2}^-] \rightarrow \Sigma_b^{*0} \pi^-}$, (d) $g_{\Xi_b^{\prime-}[{3\over2}^-] \rightarrow \Xi_b^{*0} \pi^-}$, (e) $g_{\Xi_b^{\prime-}[{3\over2}^-] \rightarrow \Sigma_b^{*0} K^-}$ and (f) $g_{\Omega_b^{-}[{3\over2}^-] \rightarrow \Xi_b^{*0} K^-}$ as functions of the Borel mass $T$. Here the excited baryons $\Sigma_b^-$, $\Xi^{\prime-}_b$, and $\Omega^-_b$ belong to the baryon doublet $[\mathbf{6}_F, 1, 0, \rho]$.
\label{fig:610rho}}
\end{figure}

The baryon doublet $[\mathbf{6}_F, 1, 0, \rho]$ contains six bottom baryons, including $\Sigma_b({1\over2}^-/{3\over2}^-)$, $\Xi^\prime_b({1\over2}^-/{3\over2}^-)$ and $\Omega_b({1\over2}^-/{3\over2}^-)$. There are altogether six non-vanishing decay channels: $(l)$, $(o)$, $(s)$, $(t)$, $(u)$ and $(v)$. We use light-cone sum rules with HQET to separately investigate them, and the relevant coupling constants are extracted to be
\begin{eqnarray}
\nonumber &(l)& g_{\Sigma_b^-[{1\over2}^-] \rightarrow \Sigma_b^{0} \pi^-} = 3.41~{^{+1.74}_{-1.33}} \, ,
\\
\nonumber &(o)& g_{\Xi_b^{\prime-}[{1\over2}^-] \rightarrow \Xi_b^{\prime0}\pi^-} = 2.31~{^{+1.13}_{-0.87}} \, ,
\\
&(s)& g_{\Sigma_b^{-}[{3\over2}^-] \rightarrow \Sigma_b^{*0} \pi^-} = 2.28~{^{+1.16}_{-0.89}} \, ,
\\
\nonumber &(t)& g_{\Xi_b^{\prime-}[{3\over2}^-] \rightarrow \Xi_b^{*0} \pi^-} = 1.54~{^{+0.75}_{-0.58}} \, ,
\\
\nonumber &(u)& g_{\Xi_b^{\prime-}[{3\over2}^-] \rightarrow \Sigma_b^{*0} K^-} = 2.10~{^{+1.07}_{-0.79}} \, ,
\\
\nonumber &(v)& g_{\Omega_b^{-}[{3\over2}^-] \rightarrow \Xi_b^{*0} K^-} = 2.72~{^{+1.29}_{-0.96}} \, .
\end{eqnarray}
For completeness, we show them as functions of the Borel mass $T$ in Fig.~\ref{fig:610rho}.
Using the above values, we can further extract their decay widths to be
\begin{eqnarray}
\nonumber &(l)& \Gamma_{\Sigma_b[{1\over2}^-] \rightarrow \Sigma_b \pi} = 850~{^{+1100}_{-540}} {\rm~MeV} \, ,
\\
\nonumber &(o)& \Gamma_{\Xi_b^{\prime}[{1\over2}^-] \rightarrow \Xi_b^{\prime}\pi} = 310~{^{+370}_{-190}} {\rm~MeV} \, ,
\\
&(s)& \Gamma_{\Sigma_b[{3\over2}^-] \rightarrow \Sigma_b^{*} \pi} = 350~{^{+440}_{-220}} {\rm~MeV} \, ,
\\
\nonumber &(t)& \Gamma_{\Xi_b^{\prime}[{3\over2}^-] \rightarrow \Xi_b^{*} \pi} = 130~{^{+150}_{-80}} {\rm~MeV} \, ,
\\
\nonumber &(u)& \Gamma_{\Xi_b^{\prime}[{3\over2}^-] \rightarrow \Sigma_b^{*} K \rightarrow \Lambda_b \pi K} = 0.029~{^{+0.036}_{-0.017}} {\rm~MeV} \, ,
\\
\nonumber &(v)& \Gamma_{\Omega_b[{3\over2}^-] \rightarrow \Xi_b^{*} K} \rightarrow {~\rm kinematically~forbidden} \, .
\end{eqnarray}
The $S$-wave decay channel (v) $\Omega_b({3/2}^-) \to \Xi_b^*({3/2}^+) K$ for the $\Omega_b({3/2}^-)$ belonging to $[\mathbf{6}_F,1,0,\rho]$ is kinematically forbidden when using the mass value $M_{\Omega_b({3/2}^-)} = 6.42 \pm 0.11$~GeV listed in Table~\ref{tab:pwave}, while the same channel for the $\Omega_b({3/2}^-)$ belonging to $[\mathbf{6}_F,2,1,\lambda]$ is kinematically allowed when using the other value $M_{\Omega_b({3/2}^-)} = 6.46 \pm 0.12$~GeV, although the obtained decay width is quite small.

\subsection{The baryon singlet $[\mathbf{6}_F, 0, 1, \lambda]$}

\begin{figure}[htb]
\begin{center}
\subfigure[]{
\scalebox{0.6}{\includegraphics{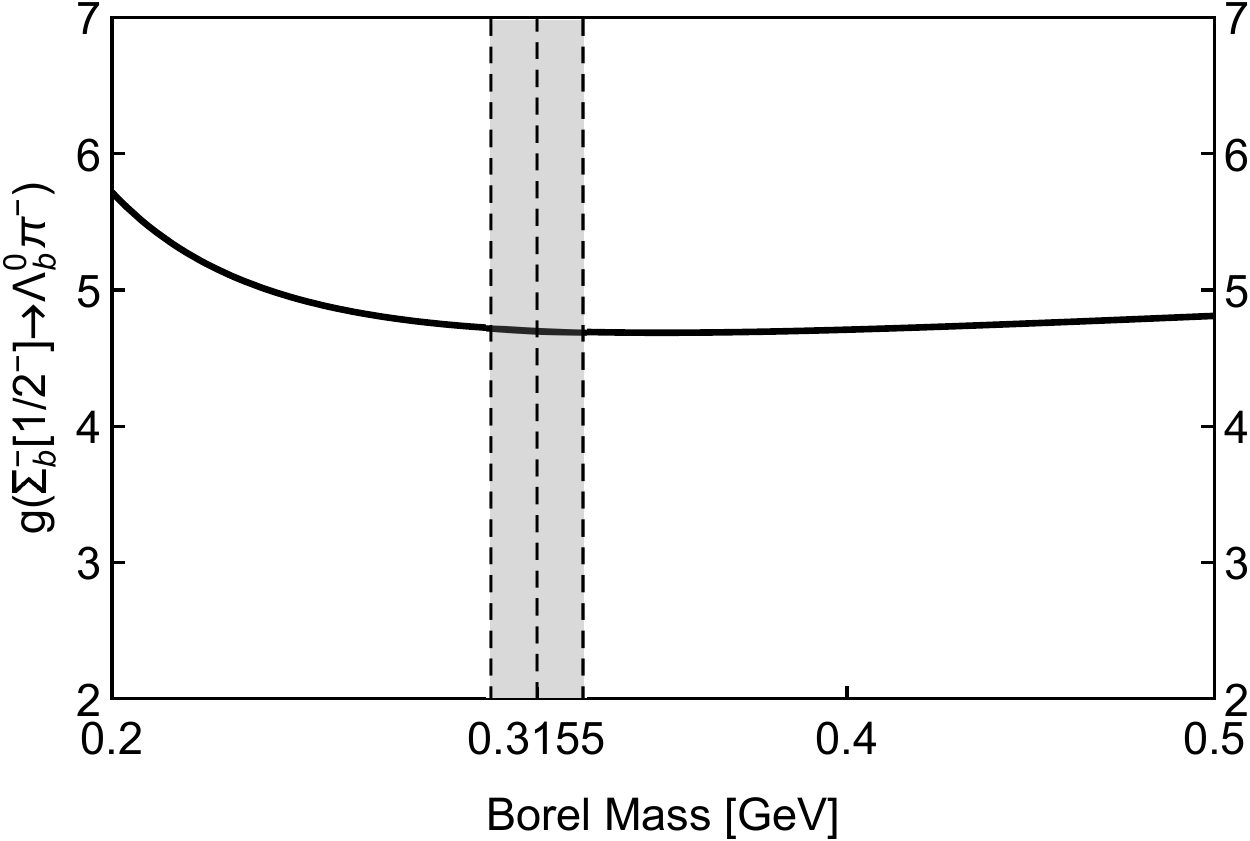}}}
~~~~~
\subfigure[]{
\scalebox{0.6}{\includegraphics{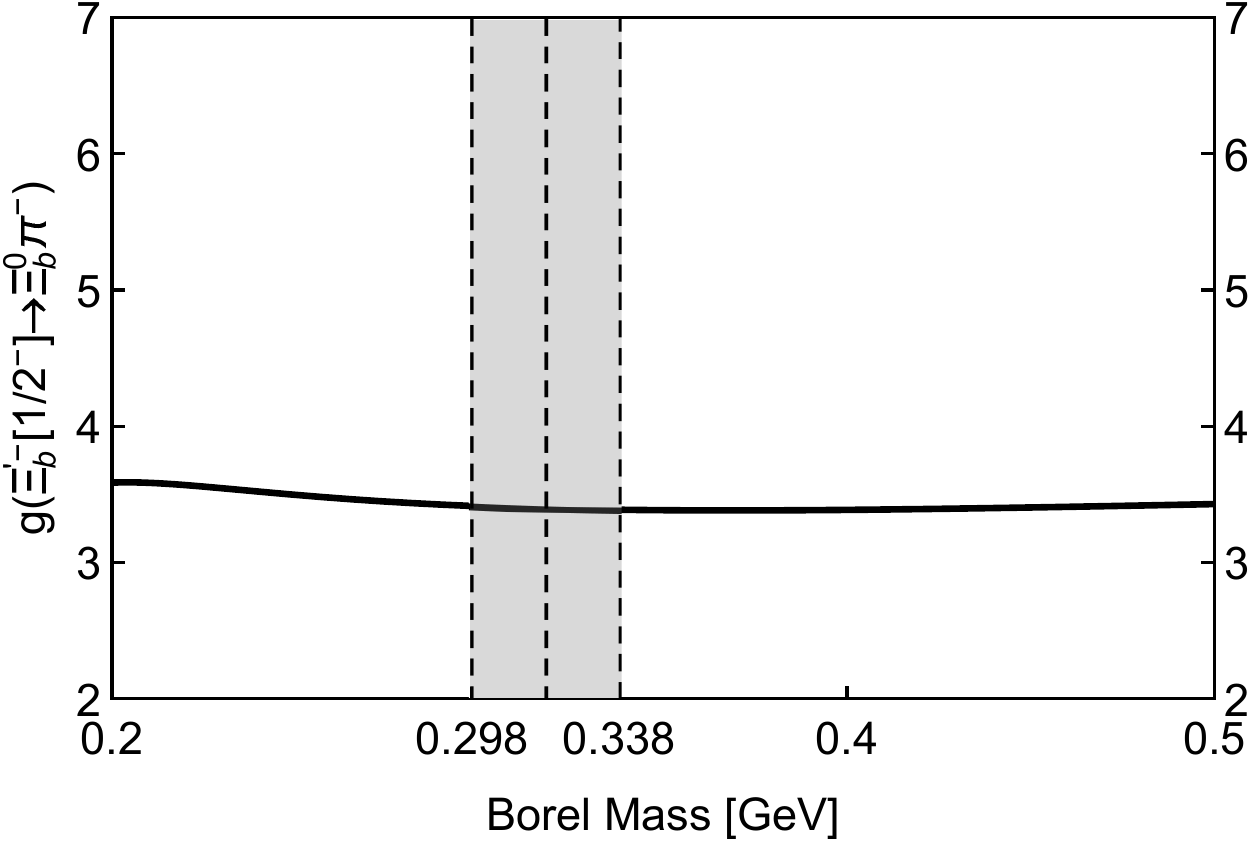}}}
\\
\subfigure[]{
\scalebox{0.6}{\includegraphics{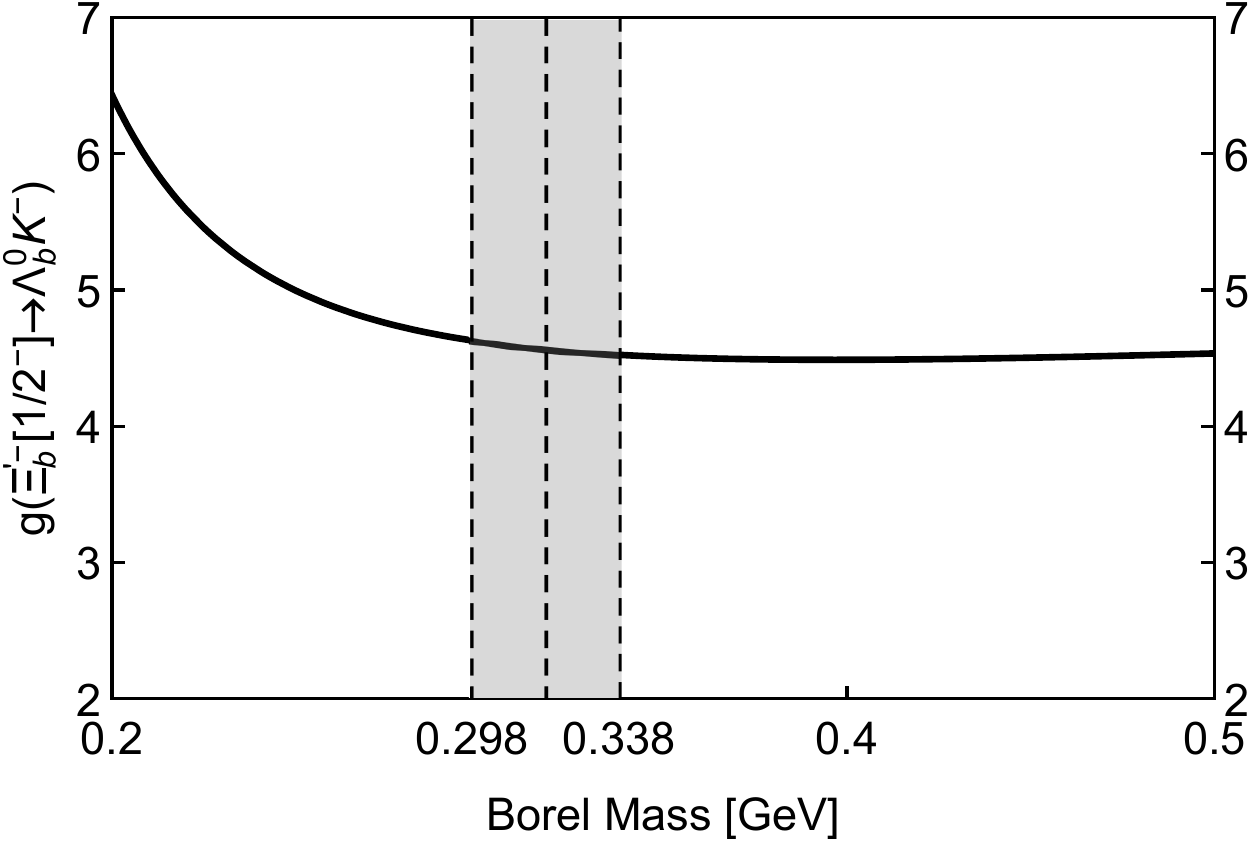}}}
~~~~~
\subfigure[]{
\scalebox{0.6}{\includegraphics{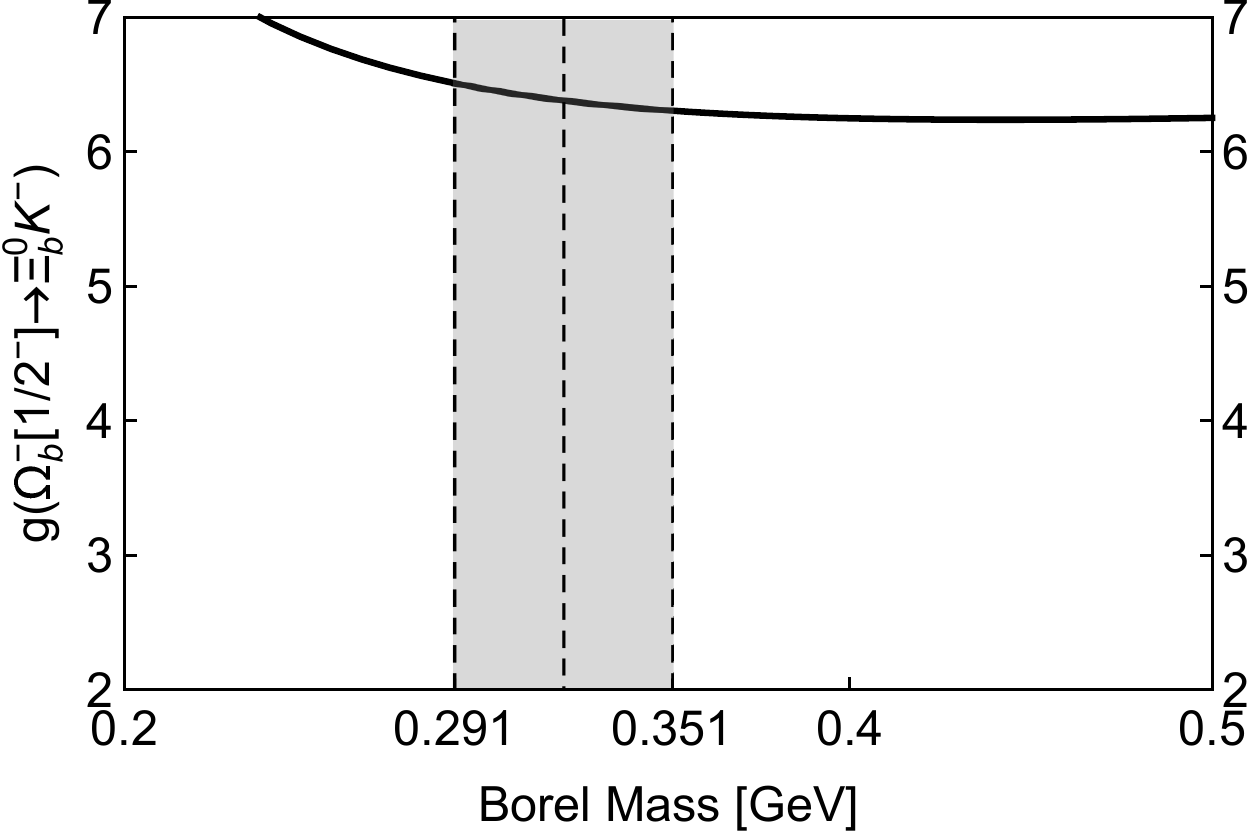}}}
\end{center}
\caption{The $S$-wave decay coupling constants (a) $g_{\Sigma_b^-[{1\over2}^-] \rightarrow \Lambda_b^{0} \pi^-}$, (b) $g_{\Xi_b^{\prime-}[{1\over2}^-] \rightarrow \Xi_b^{0}\pi^-}$, (c) $g_{\Xi_b^{\prime-}[{1\over2}^-] \rightarrow \Lambda_b^{0} K^-}$ and (d) $g_{\Omega_b^{-}[{1\over2}^-] \rightarrow \Xi_b^{0} K^-}$ as functions of the Borel mass $T$. Here the excited baryons $\Sigma_b^-$, $\Xi^{\prime-}_b$, and $\Omega^-_b$ belong to the baryon singlet $[\mathbf{6}_F, 0, 1, \lambda]$.
\label{fig:601lambda}}
\end{figure}

The baryon singlet $[\mathbf{6}_F, 0, 1, \lambda]$ contains three bottom baryons, including $\Sigma_b({1\over2}^-)$, $\Xi^\prime_b({1\over2}^-)$ and $\Omega_b({1\over2}^-)$. There are altogether four non-vanishing decay channels: $(k)$, $(m)$, $(n)$ and $(q)$. We use light-cone sum rules with HQET to separately investigate them, and the relevant coupling constants are extracted to be
\begin{eqnarray}
\nonumber &(k)& g_{\Sigma_b^-[{1\over2}^-] \rightarrow \Lambda_b^{0} \pi^-} = 4.70~{^{+2.39}_{-1.84}} \, ,
\\
&(m)& g_{\Xi_b^{\prime-}[{1\over2}^-] \rightarrow \Xi_b^{0}\pi^-} = 3.40~{^{+1.69}_{-1.30}} \, ,
\\
\nonumber &(n)& g_{\Xi_b^{\prime-}[{1\over2}^-] \rightarrow \Lambda_b^{0} K^-} = 4.56~{^{+2.35}_{-1.74}} \, ,
\\
\nonumber &(q)& g_{\Omega_b^{-}[{1\over2}^-] \rightarrow \Xi_b^{0} K^-} = 6.38~{^{+3.16}_{-2.35}} \, .
\end{eqnarray}
For completeness, we show them as functions of the Borel mass $T$ in Fig.~\ref{fig:601lambda}.
Using the above values, we can further extract their decay widths to be
\begin{eqnarray}
\nonumber &(k)& \Gamma_{\Sigma_b[{1\over2}^-] \rightarrow \Lambda_b \pi} = 1400~{^{+1800}_{-900}} {\rm~MeV} \, ,
\\
&(m)& \Gamma_{\Xi_b^{\prime}[{1\over2}^-] \rightarrow \Xi_b\pi} = 1000~{^{+1300}_{-630}} {\rm~MeV} \, ,
\\
\nonumber &(n)& \Gamma_{\Xi_b^{\prime}[{1\over2}^-] \rightarrow \Lambda_b K} = 1000~{^{+1300}_{-620}} {\rm~MeV} \, ,
\\
\nonumber &(r)& \Gamma_{\Omega_b[{1\over2}^-] \rightarrow \Xi_b K} = 3900~{^{+4900}_{-2400}} {\rm~MeV} \, .
\end{eqnarray}

\subsection{The baryon doublet $[\mathbf{6}_F, 2, 1, \lambda]$}

\begin{figure}[htb]
\begin{center}
\subfigure[]{
\scalebox{0.6}{\includegraphics{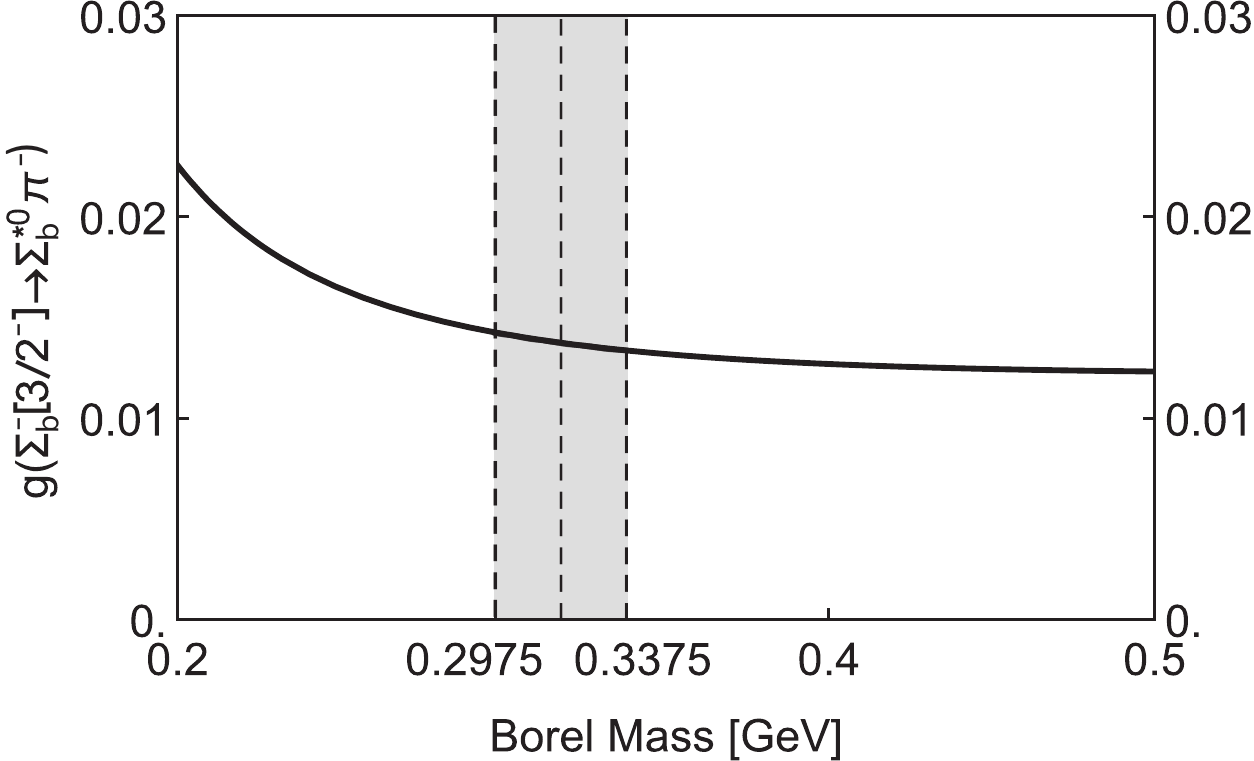}}}
~~~~~
\subfigure[]{
\scalebox{0.6}{\includegraphics{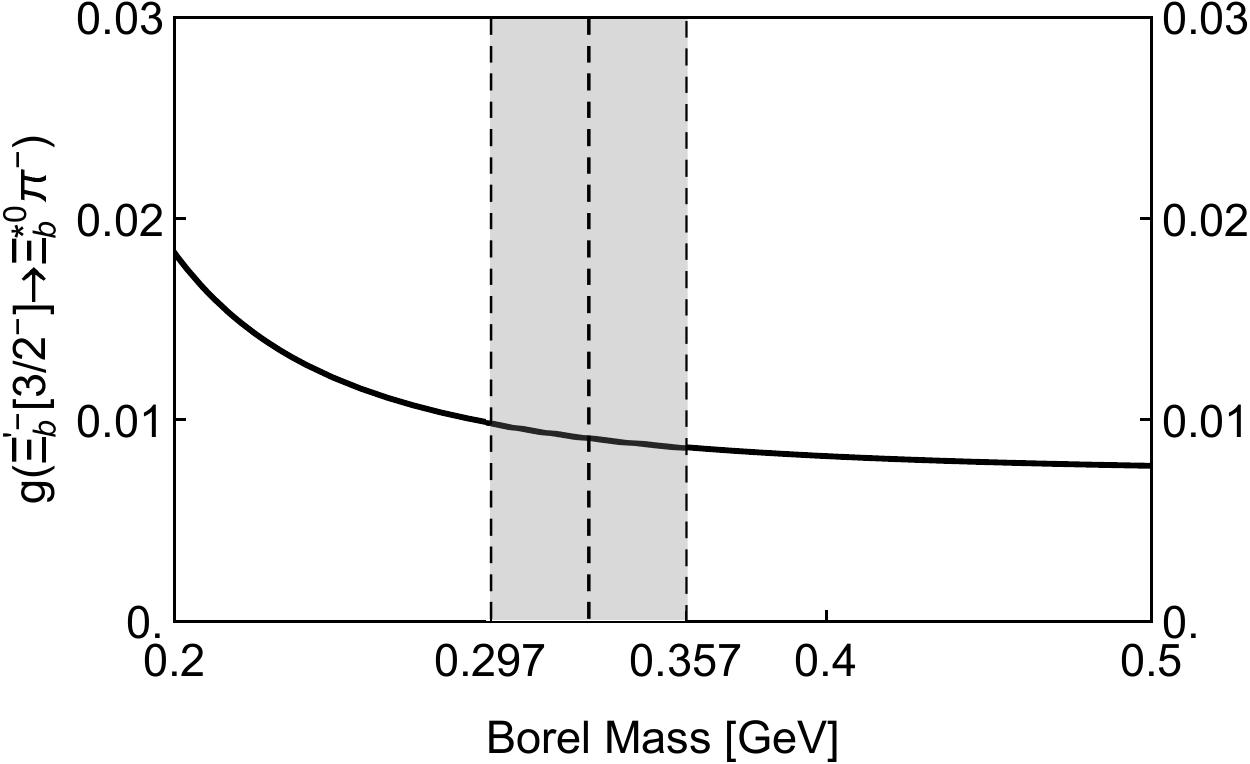}}}
\\
\subfigure[]{
\scalebox{0.6}{\includegraphics{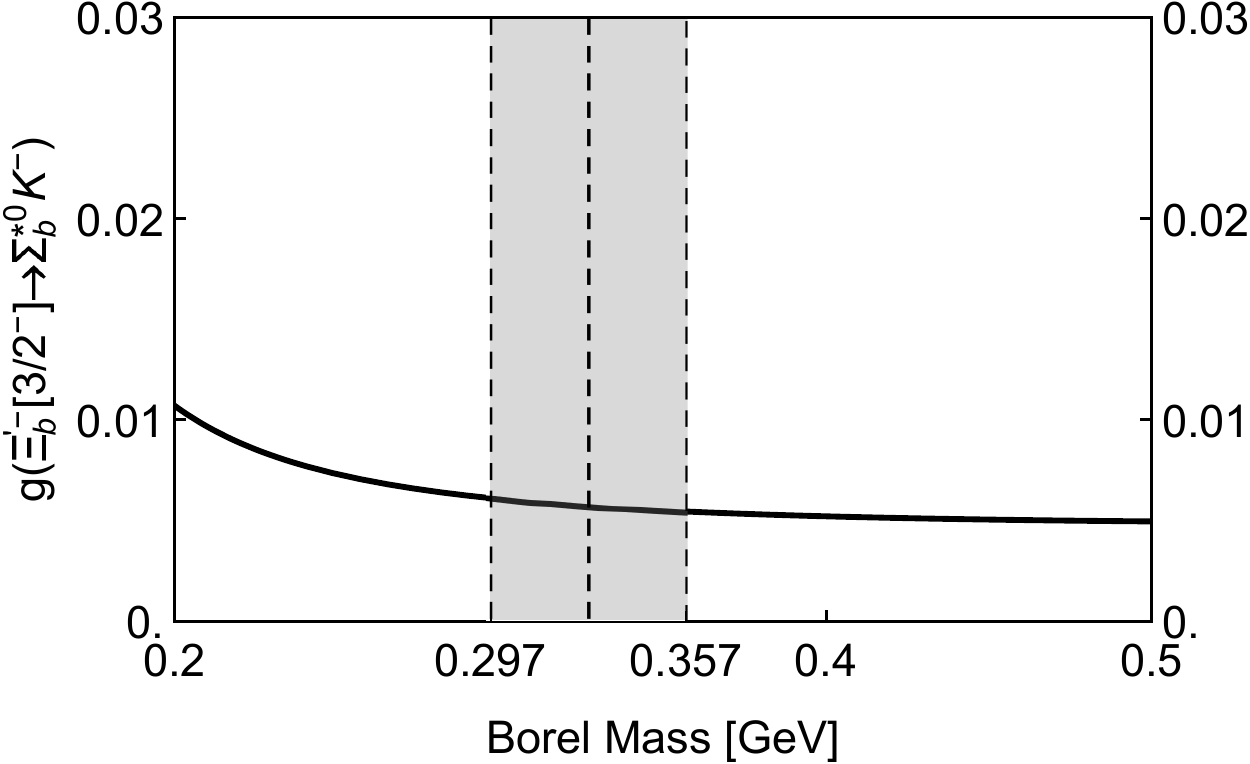}}}
~~~~~
\subfigure[]{
\scalebox{0.6}{\includegraphics{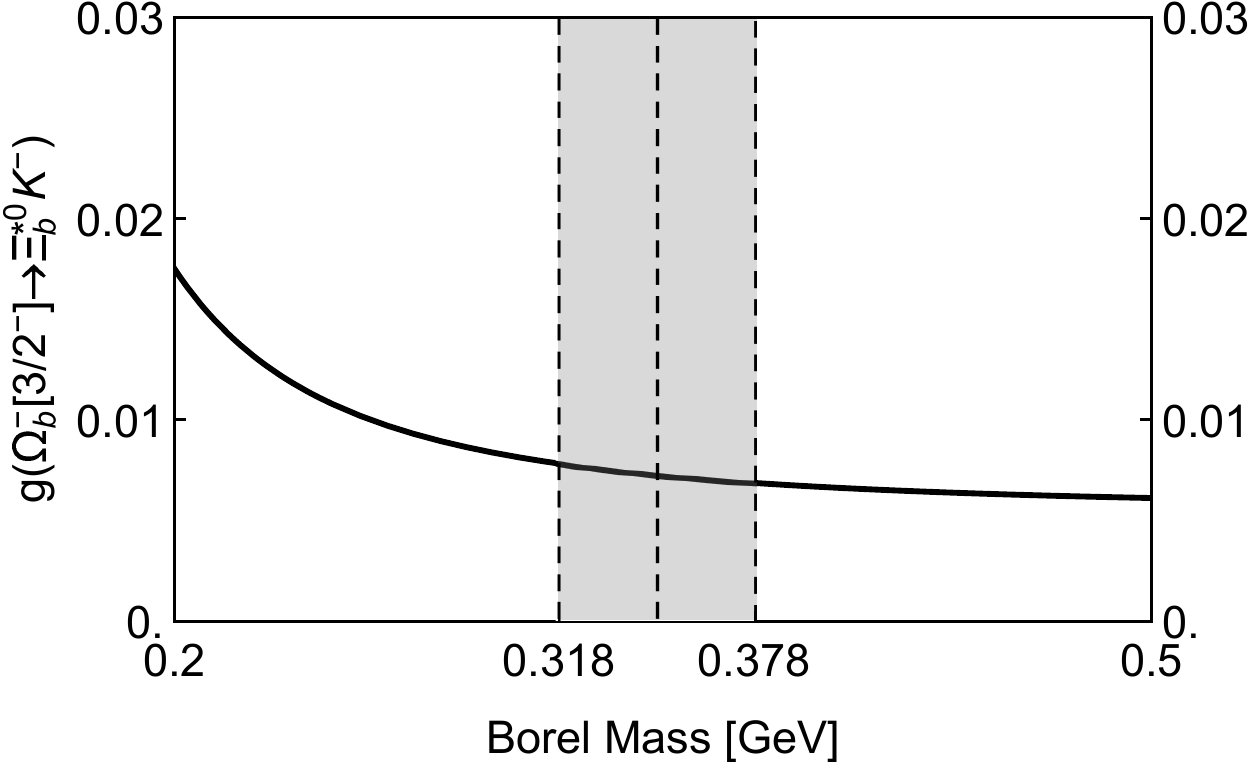}}}
\end{center}
\caption{The $S$-wave decay coupling constants (a) $g_{\Sigma_b^-[{3\over2}^-] \rightarrow \Sigma_b^{*0} \pi^-}$, (b) $g_{\Xi_b^{\prime-}[{3\over2}^-] \rightarrow \Xi_b^{*0}\pi^-}$, (c) $g_{\Xi_b^{\prime-}[{3\over2}^-] \rightarrow \Sigma_b^{*0} K^-}$ and (d) $g_{\Omega_b^{-}[{3\over2}^-] \rightarrow \Xi_b^{*0} K^-}$ as functions of the Borel mass $T$. Here the excited baryons $\Sigma_b^-$, $\Xi^{\prime-}_b$, and $\Omega^-_b$ belong to the baryon doublet $[\mathbf{6}_F, 2, 1, \lambda]$.
\label{fig:621lambda}}
\end{figure}

The baryon doublet $[\mathbf{6}_F, 2, 1, \lambda]$ contains six bottom baryons, including $\Sigma_b({3\over2}^-/{5\over2}^-)$, $\Xi^\prime_b({3\over2}^-/{5\over2}^-)$ and $\Omega_b({3\over2}^-/{5\over2}^-)$. There are altogether four non-vanishing decay channels: $(s)$, $(t)$, $(u)$ and $(v)$. We use light-cone sum rules with HQET to separately investigate them, and the relevant coupling constants are extracted to be
\begin{eqnarray}
\nonumber &(s)& g_{\Sigma_b^{-}[{3\over2}^-] \rightarrow \Sigma_b^{*0} \pi^-} = 0.014~{^{+0.008}_{-0.007}} \, ,
\\
&(t)& g_{\Xi_b^{\prime-}[{3\over2}^-] \rightarrow \Xi_b^{*0} \pi^-} = 0.009~{^{+0.005}_{-0.005}} \, ,
\\
\nonumber &(u)& g_{\Xi_b^{\prime-}[{3\over2}^-] \rightarrow \Sigma_b^{*0} K^-} = 0.006~{^{+0.010}_{-0.006}} \, ,
\\
\nonumber &(v)& g_{\Omega_b^{-}[{3\over2}^-] \rightarrow \Xi_b^{*0} K^-} = 0.007~{^{+0.012}_{-0.007}} \, .
\end{eqnarray}
For completeness, we show them as functions of the Borel mass $T$ in Fig.~\ref{fig:621lambda}.
Using the above values, we can further extract their decay widths to be
\begin{eqnarray}
\nonumber &(s)& \Gamma_{\Sigma_b[{3\over2}^-] \rightarrow \Sigma_b^{*} \pi} = 0.013~{^{+0.019}_{-0.010}} {\rm~MeV} \, ,
\\
&(t)& \Gamma_{\Xi_b^{\prime}[{3\over2}^-] \rightarrow \Xi_b^{*} \pi} = 0.004~{^{+0.006}_{-0.003}} {\rm~MeV} \, ,
\\
\nonumber &(u)& \Gamma_{\Xi_b^{\prime}[{3\over2}^-] \rightarrow \Sigma_b^{*} K \rightarrow \Lambda_b \pi K} = 2~{^{+14}_{-2}} \times 10^{-7} {\rm~MeV} \, ,
\\
\nonumber &(v)& \Gamma_{\Omega_b[{3\over2}^-] \rightarrow \Xi_b^{*} K} = 0.001~{^{+0.008}_{-0.001}} {\rm~MeV} \, .
\end{eqnarray}
The large uncertainties of the above results mainly come from the light-cone distribution amplitudes of pseudoscalar mesons.

%
\section{$D$-wave decay properties of $P$-wave bottom baryons}\label{sec:ddecay}
%

In the previous section we studied the $S$-wave decay properties of $P$-wave bottom baryons, and the results are summarized in Table~\ref{tab:sdecay}. From this table we find that none of the three bottom baryon multiplets, $[\mathbf{6}_F, 0, 1, \lambda]$, $[\mathbf{6}_F, 1, 0, \rho]$ and $[\mathbf{6}_F, 2, 1, \lambda]$, can explain the decay widths of the $\Sigma_{b}(6097)^\pm$ and $\Xi_{b}(6227)^{-}$~\cite{Aaij:2018yqz,Aaij:2018tnn}. This forces us to further investigate their $D$-wave decay properties. We note that another opinion is to study the mixing of the above baryon multiplets, which can happen because the heavy quark effective theory is not perfect, but we shall not consider this in the present study.

\begin{figure}[htb]
\begin{center}
\subfigure[]{
\scalebox{0.6}{\includegraphics{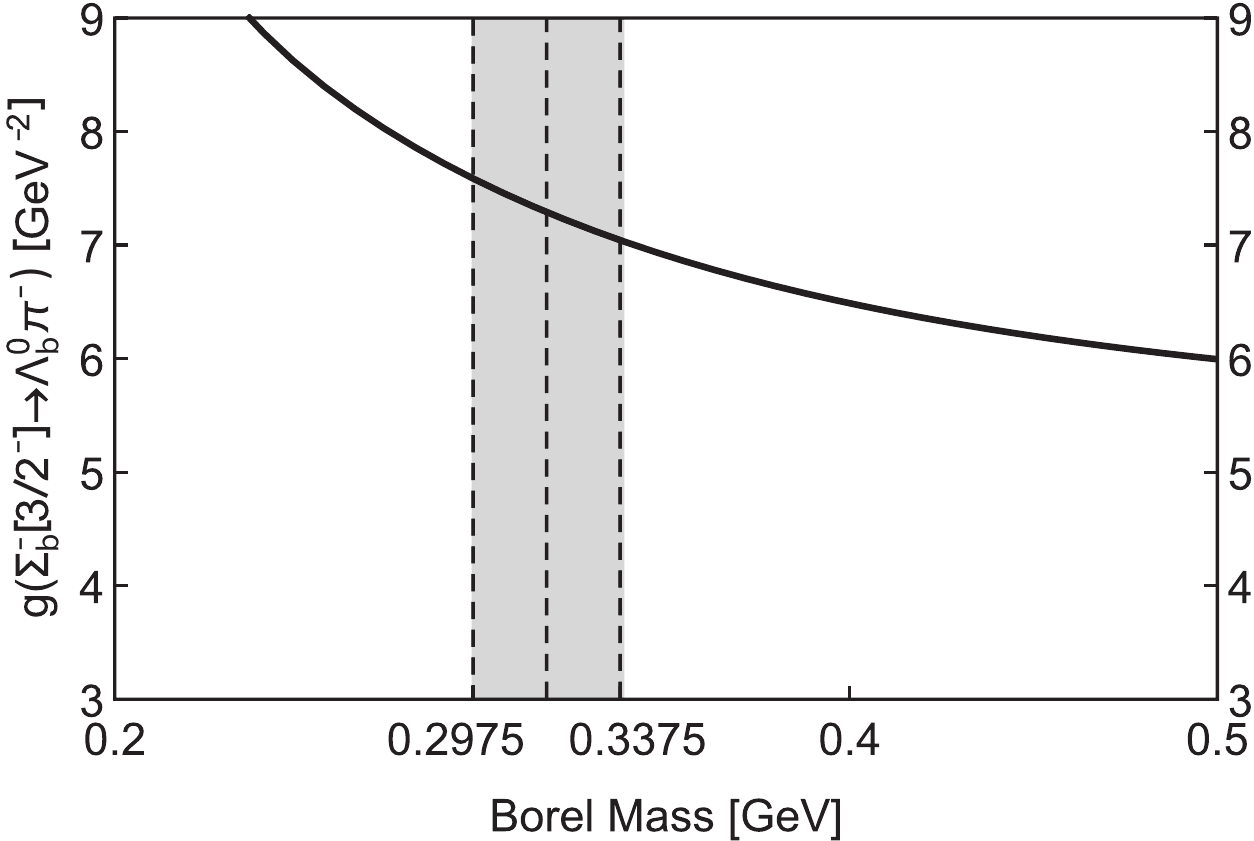}}}
~~~~~
\subfigure[]{
\scalebox{0.6}{\includegraphics{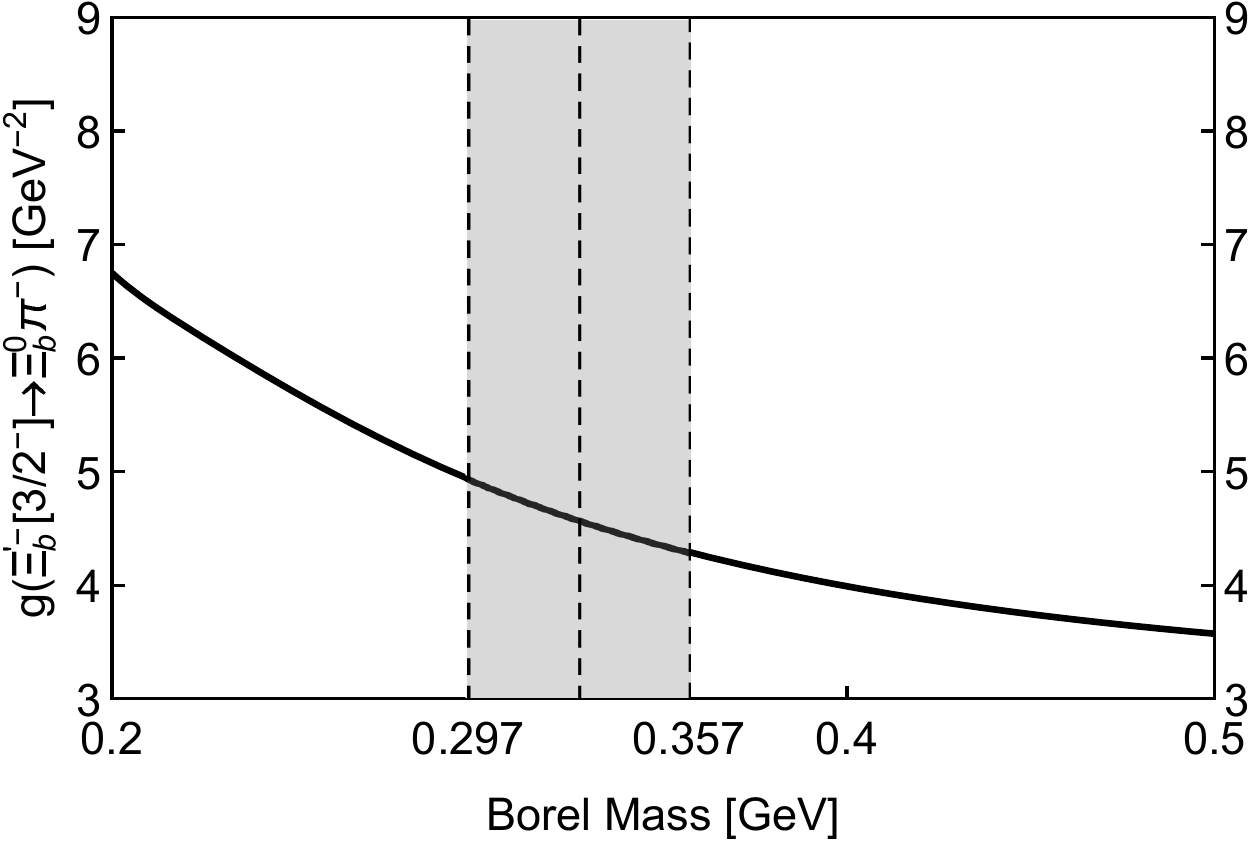}}}
\\
\subfigure[]{
\scalebox{0.6}{\includegraphics{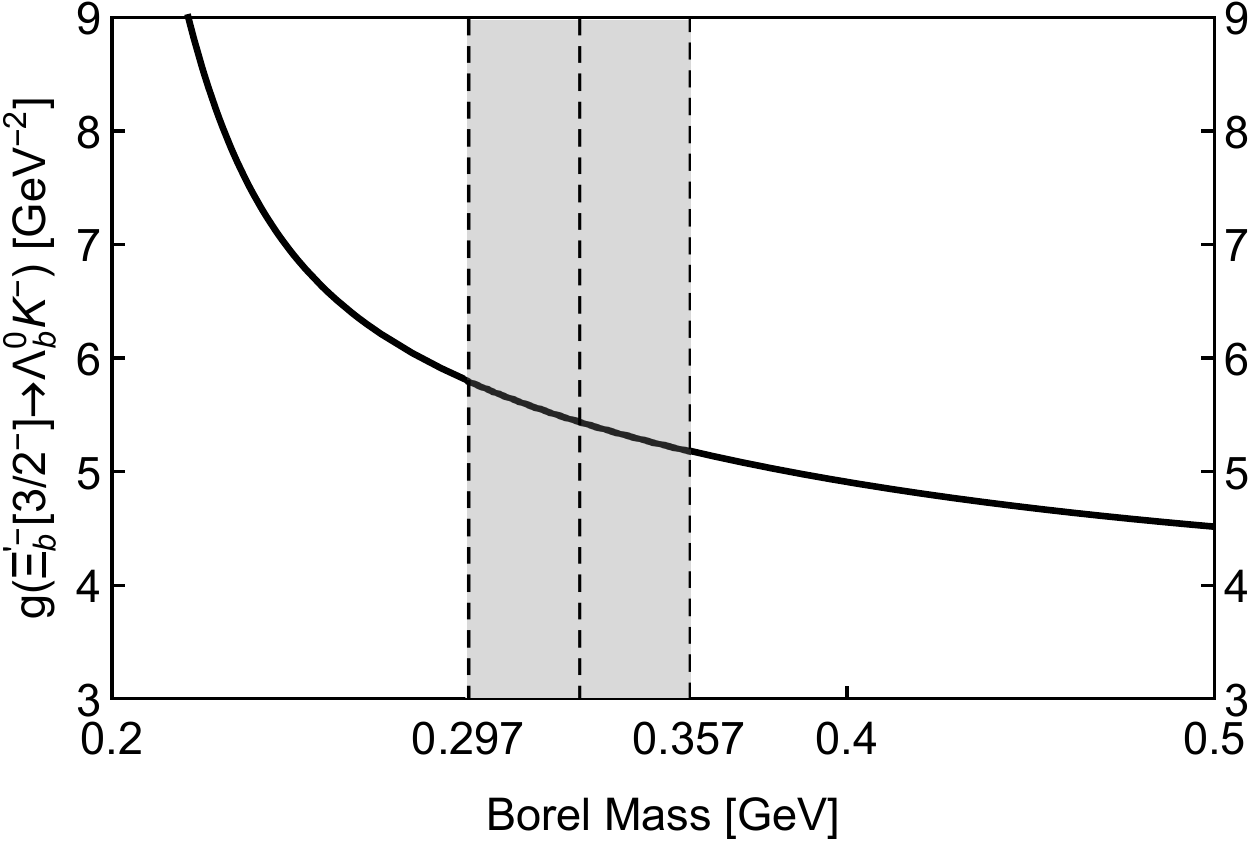}}}
~~~~~
\subfigure[]{
\scalebox{0.6}{\includegraphics{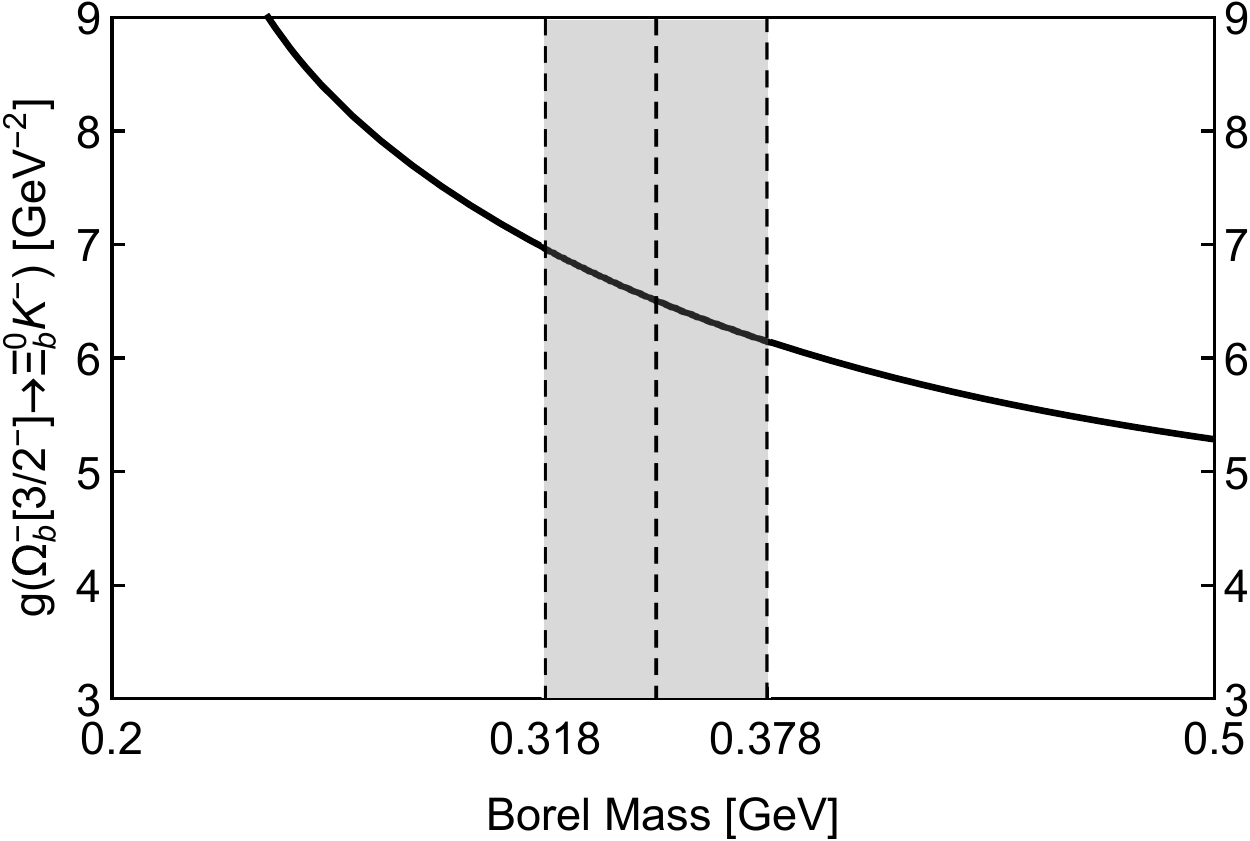}}}
\end{center}
\caption{The $D$-wave decay coupling constants (a) $g_{\Sigma_b^-[{3\over2}^-] \rightarrow \Lambda_b^{0} \pi^-}$, (b) $g_{\Xi_b^{\prime-}[{3\over2}^-] \rightarrow \Xi_b^{0}\pi^-}$, (c) $g_{\Xi_b^{\prime-}[{3\over2}^-] \rightarrow \Lambda_b^{0} K^-}$ and (d) $g_{\Omega_b^{-}[{3\over2}^-] \rightarrow \Xi_b^{0} K^-}$ as functions of the Borel mass $T$. Here the excited baryons $\Sigma_b^-$, $\Xi^{\prime-}_b$, and $\Omega^-_b$ belong to the baryon doublet $[\mathbf{6}_F, 2, 1, \lambda]$.
\label{fig:621lambda2}}
\end{figure}

Because the widths of the bottom baryons belonging to the baryon multiplets $[\mathbf{6}_F, 0, 1, \lambda]$ and $[\mathbf{6}_F, 1, 0, \rho]$ are already too large, we only need to consider the last multiplet $[\mathbf{6}_F, 2, 1, \lambda]$. Especially, in the present study we study their $D$-wave decays into the lowest-lying bottom baryons of flavor $\mathbf{\bar 3}_F$ accompanied by a pseudoscalar meson ($\pi$ or $K$), {\it i.e.},
\begin{eqnarray}
&(w)& {\bf \Gamma\Big[} \Sigma_b(3/2^-) \rightarrow \Lambda_b(1/2^+) + \pi {\Big ]}
= {\bf \Gamma\Big[} \Sigma_b^{-}(3/2^-) \rightarrow \Lambda_b^{0}(1/2^+) + \pi^- {\Big ]} \, ,
\\ &(x)& {\bf \Gamma\Big[}\Xi_b^{\prime}(3/2^-) \rightarrow \Xi_b(1/2^+) + \pi{\Big ]}
= {3\over2} \times {\bf \Gamma\Big[}\Xi_b^{\prime -}(3/2^-) \rightarrow \Xi_b^{0}(1/2^+) + \pi^-{\Big ]} \, ,
\\ &(y)& {\bf \Gamma\Big[} \Xi_b^{\prime}(3/2^-) \rightarrow \Lambda_b(1/2^+) + K {\Big ]}
= {\bf \Gamma\Big[} \Xi_b^{\prime -}(3/2^-) \rightarrow \Lambda_b^{0}(1/2^+) + K^- {\Big ]} \, ,
\\ &(z)& {\bf \Gamma\Big[}\Omega_b(3/2^-) \rightarrow \Xi_b(1/2^+) + K{\Big ]}
= 2 \times {\bf \Gamma\Big[}\Omega_b^{-}(3/2^-) \rightarrow \Xi_b^{0}(1/2^+) + K^-{\Big ]} \, ,
\\ &(w^\prime)& {\bf \Gamma\Big[} \Sigma_b(5/2^-) \rightarrow \Lambda_b(1/2^+) + \pi {\Big ]}
= {\bf \Gamma\Big[} \Sigma_b^{-}(5/2^-) \rightarrow \Lambda_b^{0}(1/2^+) + \pi^- {\Big ]} \, ,
\\ &(x^\prime)& {\bf \Gamma\Big[}\Xi_b^{\prime}(5/2^-) \rightarrow \Xi_b(1/2^+) + \pi{\Big ]}
= {3\over2} \times {\bf \Gamma\Big[}\Xi_b^{\prime -}(5/2^-) \rightarrow \Xi_b^{0}(1/2^+) + \pi^-{\Big ]} \, ,
\\ &(y^\prime)& {\bf \Gamma\Big[} \Xi_b^{\prime}(5/2^-) \rightarrow \Lambda_b(1/2^+) + K {\Big ]}
= {\bf \Gamma\Big[} \Xi_b^{\prime -}(5/2^-) \rightarrow \Lambda_b^{0}(1/2^+) + K^- {\Big ]} \, ,
\\ &(z^\prime)& {\bf \Gamma\Big[}\Omega_b(5/2^-) \rightarrow \Xi_b(1/2^+) + K{\Big ]}
= 2 \times {\bf \Gamma\Big[}\Omega_b^{-}(5/2^-) \rightarrow \Xi_b^{0}(1/2^+) + K^-{\Big ]} \, ,
\end{eqnarray}
We can calculate their decay widths through the following Lagrangians
\begin{eqnarray}
\mathcal{L}_{X_b({3/2}^-) \rightarrow Y_b({1/2}^+) P} &=& g {\bar X_b^\mu}(3/2^-) \gamma_\nu \gamma_5 Y_b(1/2^+) \partial_\mu \partial_\nu P \, ,
\\ \mathcal{L}_{X_b({5/2}^-) \rightarrow Y_b({1/2}^+) P} &=& g {\bar X_{b}^{\mu\nu}}(5/2^-) Y_{b}(1/2^+) \partial_\mu \partial_\nu P \, .
\end{eqnarray}
We use light-cone sum rules with HQET to separately investigate them, and the obtained sum rule equations are given in Appendix~\ref{sec:othersumrule}. The relevant coupling constants are extracted to be
\begin{eqnarray}
\nonumber &(w)& g_{\Sigma_b^{-}[{3\over2}^-] \rightarrow \Lambda_b^{0} \pi^-} = 7.29~{^{+3.65}_{-2.75}}~\mbox{GeV}^{-2} \, ,
\\
\nonumber&(x)& g_{\Xi_b^{\prime-}[{3\over2}^-] \rightarrow \Xi_b^{0} \pi^-} = 4.57~{^{+2.17}_{-1.67}}~\mbox{GeV}^{-2} \, ,
\\
\nonumber &(y)& g_{\Xi_b^{\prime-}[{3\over2}^-] \rightarrow \Lambda_b^{0} K^-} = 5.44~{^{+2.65}_{-1.95}}~\mbox{GeV}^{-2} \, ,
\\
&(z)& g_{\Omega_b^{-}[{3\over2}^-] \rightarrow \Xi_b^{0} K^-} = 6.51~{^{+2.97}_{-2.22}}~\mbox{GeV}^{-2} \, ,
\\
\nonumber &(w^\prime)& g_{\Sigma_b^{-}[{5\over2}^-] \rightarrow \Lambda_b^{0} \pi^-} = 0 \, ,
\\
\nonumber &(x^\prime)& g_{\Xi_b^{\prime-}[{5\over2}^-] \rightarrow \Xi_b^{0} \pi^-} = 0 \, ,
\\
\nonumber &(y^\prime)& g_{\Xi_b^{\prime-}[{5\over2}^-] \rightarrow \Lambda_b^{0} K^-} = 0 \, ,
\\
\nonumber &(z^\prime)& g_{\Omega_b^{-}[{5\over2}^-] \rightarrow \Xi_b^{0} K^-} = 0 \, .
\end{eqnarray}
For completeness, we show them as functions of the Borel mass $T$ in Fig.~\ref{fig:621lambda2}.
Using the above values, we can further extract their decay widths to be
\begin{eqnarray}
\nonumber &(w)& \Gamma_{\Sigma_b[{3\over2}^-] \rightarrow \Lambda_b \pi} = 46~{^{+58}_{-28}} {\rm~MeV} \, ,
\\
\nonumber &(x)& \Gamma_{\Xi_b^{\prime}[{3\over2}^-] \rightarrow \Xi_b \pi} = 16~{^{+19}_{-10}} {\rm~MeV} \, ,
\\
\nonumber &(y)& \Gamma_{\Xi_b^{\prime}[{3\over2}^-] \rightarrow \Lambda_b K} = 6.5~{^{+7.9}_{-3.8}} {\rm~MeV} \, ,
\\
\nonumber &(z)& \Gamma_{\Omega_b[{3\over2}^-] \rightarrow \Xi_b K} = 58~{^{+65}_{-33}} {\rm~MeV} \, ,
\\
\nonumber &(w^\prime)& \Gamma_{\Sigma_b^{-}[{5\over2}^-] \rightarrow \Lambda_b^{0} \pi^-} = 0 \, ,
\\
\nonumber &(x^\prime)& \Gamma_{\Xi_b^{\prime-}[{5\over2}^-] \rightarrow \Xi_b^{0} \pi^-} = 0 \, ,
\\
\nonumber &(y^\prime)& \Gamma_{\Xi_b^{\prime-}[{5\over2}^-] \rightarrow \Lambda_b^{0} K^-} = 0 \, ,
\\
\nonumber &(z^\prime)& \Gamma_{\Omega_b^{-}[{5\over2}^-] \rightarrow \Xi_b^{0} K^-} = 0 \, .
\end{eqnarray}
The above results are summarized in Table~\ref{tab:ddecay}.

\begin{table}[hbt]
\begin{center}
\renewcommand{\arraystretch}{1.5}
\caption{$D$-wave decay properties of the $P$-wave bottom baryons belonging to the baryon doublet $[\mathbf{6}_F, 2, 1, \lambda]$.}
\begin{tabular}{c | l | c | c}
\hline\hline
Multiplets & $D$-wave decay channels & ~$g$ ($\mbox{GeV}^{-2}$)~ & $D$-wave decay width (MeV)
\\ \hline\hline
\multirow{4}{*}{$[\mathbf{6}_F,2,1,\lambda]$}    & (w) $\Sigma_b({3\over2}^-)\to \Lambda_b({1\over2}^+) \pi$ & $7.29^{+3.65}_{-2.75}$ & $ 46^{+58}_{-28}$
\\
                                                 & (x) $\Xi_b^{\prime}({3\over2}^-)\to \Xi_b({1\over2}^+) \pi$ & $4.57^{+2.17}_{-1.67}$ &$ 16^{+19}_{-10}$
\\
                                                 & (y) $\Xi_b^{\prime}({3\over2}^-)\to \Lambda_b({1\over2}^+) K$ & $5.44^{+2.65}_{-1.95}$ & $6.5^{+7.9}_{-3.8}$
\\
                                                 & (z) $\Omega_b({3\over2}^-)\to \Xi_b({1\over2}^+) K$ & $6.51^{+2.97}_{-2.22}$ & $ 58^{+65}_{-33}$
\\ \hline \hline
\end{tabular}
\label{tab:ddecay}
\end{center}
\end{table}

%
\section{Summary and Discussions}\label{sec:summary}
%

In this paper we have investigated the $\Xi_{b}(6227)^{-}$ and $\Sigma_{b}(6097)^{\pm}$ newly observed by LHCb~\cite{Aaij:2018yqz,Aaij:2018tnn}. We use the method of QCD sum rules within the framework of heavy quark effective theory to study their mass spectrum, and use the method of light-cone sum rules still within the heavy quark effective theory to study their decay properties. Based on our previous studies~\cite{Chen:2015kpa,Mao:2015gya,Chen:2017sci}, we have investigated three $P$-wave bottom baryon multiplets, $[\mathbf{6}_F, 0, 1, \lambda]$, $[\mathbf{6}_F, 1, 0, \rho]$ and $[\mathbf{6}_F, 2, 1, \lambda]$, and their masses and decay widths are extracted and summarized in Tables~\ref{tab:pwave},~\ref{tab:sdecay}, and~\ref{tab:ddecay}. Especially, the masses and decay widths of the $\Sigma_b(3/2^-)$ and $\Xi^\prime_b(3/2^-)$ belonging to $[\mathbf{6}_F, 2, 1, \lambda]$ are extracted to be
\begin{eqnarray*}
M_{\Sigma_b(3/2^-)} &=& 6.10 \pm 0.12 {~\rm GeV} \, ,
\\ \Gamma_{\Sigma_b(3/2^-)} &=& 46~{^{+58}_{-28}} {~\rm MeV}~({\rm total}) \, ,
\\ M_{\Xi^\prime_b(3/2^-)} &=& 6.27 \pm 0.12 {~\rm GeV} \, ,
\\ \Gamma_{\Xi^\prime_b(3/2^-)} &=& 23~{^{+27}_{-14}} {~\rm MeV}~({\rm total}) \, ,
\end{eqnarray*}
which are consistent with the experimental parameters of the $\Xi_{b}(6227)^{-}$ and $\Sigma_{b}(6097)^{\pm}$~\cite{Aaij:2018yqz,Aaij:2018tnn}.
Their non-vanishing decay channels are extracted to be
\begin{eqnarray*}
\Gamma_{\Sigma_b(3/2^-) \rightarrow \Lambda_b \pi} &=& 46~{^{+58}_{-28}}{~\rm MeV} \, ,
\\ \Gamma_{\Sigma_b(3/2^-) \rightarrow \Sigma_b^* \pi} &=& 1.3~{^{+1.9}_{-1.0}}\times 10^{-2} {~\rm MeV} \, ,
\\ \Gamma_{\Xi^\prime_b(3/2^-) \rightarrow \Xi_b \pi} &=& 16~{^{+19}_{-10}}{~\rm MeV} \, ,
\\ \Gamma_{\Xi^\prime_b(3/2^-) \rightarrow \Lambda_b K} &=& 6.5~{^{+7.9}_{-3.8}} {~\rm MeV}\, ,
\\ \Gamma_{\Xi^\prime_b(3/2^-) \rightarrow \Xi^*_b \pi} &=& 4~{^{+6}_{-3}}\times 10^{-3} {~\rm MeV} \, ,
\\ \Gamma_{\Xi^\prime_b(3/2^-) \rightarrow \Sigma_b^{*} K} &=& 2~{^{+14}_{-2}} \times 10^{-7} {\rm~MeV} \, .
\end{eqnarray*}
Especially, the branching ratio
\begin{eqnarray*}
{{\mathcal B}(\Sigma_b(3/2^-)^- \rightarrow \Lambda_b^0 K^-) \over {\mathcal B}(\Sigma_b(3/2^-)^- \rightarrow \Xi_b^0 \pi^-)} = 0.6~{^{+1.1}_{-0.5}} \, ,
\end{eqnarray*}
is also consistent with the one measured by LHCb~\cite{Aaij:2018yqz} (see Eq.~(\ref{eq:br})). Its uncertainty come from the Borel mass, the parameters of the $\Lambda_b^{0}$, the parameters of the $\Xi_b^{0}$, the parameters of the $\Sigma_b^-[{3\over2}^-]$, and various quark masses and condensates listed in Eq.~(\ref{eq:condensates}).

Summarizing the above results, we conclude that the $\Sigma_{b}(6097)^\pm$ and $\Xi_{b}(6227)^{-}$ can be well interpreted as $P$-wave bottom baryons with the spin-parity $J^P = 3/2^-$, which belong to the baryon doublet $[\mathbf{6}_F, 2, 1, \lambda]$. We predict the mass and decay width of their $\Omega_b(3/2^-)$ partner state to be
\begin{eqnarray}
M_{\Omega_b(3/2^-)} &=& 6.46~ \pm 0.12 {~\rm GeV} \, ,
\\ \Gamma_{\Omega_b(3/2^-)} &=& 58~{^{+65}_{-33}} {~\rm MeV} \, ,
\end{eqnarray}
with the following decay channels
\begin{eqnarray}
\Gamma_{\Omega_b(3/2^-) \rightarrow \Xi_b K} &=& 58~{^{+65}_{-33}} {~\rm MeV} \, ,
\\ \Gamma_{\Omega_b(3/2^-) \rightarrow \Xi_b^* K} &=& 1~{^{+8}_{-1}} \times 10^{-3} {~\rm MeV} \, .
\end{eqnarray}
Moreover, the baryon doublet $[\mathbf{6}_F, 2, 1, \lambda]$ contains six bottom baryons: $\Sigma_b({3\over2}^-/{5\over2}^-)$, $\Xi^\prime_b({3\over2}^-/{5\over2}^-)$ and $\Omega_b({3\over2}^-/{5\over2}^-)$. We find that the three bottom baryons of $J^P = 5/2^-$ have quite narrow widths, and their masses as well as the mass differences within the same doublet are extracted to be
\begin{eqnarray*}
M_{\Sigma_b(5/2^-)} &=& 6.11 \pm 0.12~{\rm GeV} \, , \, M_{\Sigma_b(5/2^-)} -  M_{\Sigma_b(3/2^-)} = 13 \pm 5~{\rm MeV} \, ,
\\M_{\Xi_b^{\prime}(5/2^-)} &=& 6.29 \pm 0.11~{\rm GeV} \, , \, M_{\Xi^\prime_b(5/2^-)} -  M_{\Xi^\prime_b(3/2^-)} = 12 \pm 5~{\rm MeV} \, ,
\\M_{\Omega_b(5/2^-)} &=& 6.47 \pm 0.12~{\rm GeV} \, , \,  M_{\Omega_b(5/2^-)} -  M_{\Omega_b(3/2^-)} = 11 \pm 5~{\rm MeV} \, .
\end{eqnarray*}
We propose to search for the above four $P$-wave bottom baryons in further LHCb experiments.

%
\section*{Acknowledgments}

E. L. C is grateful to A. Hosaka and all group members for their considerable help in life and
meaningful discussions on work during his stay at RCNP, Osaka University.
This project is supported by
the National Natural Science Foundation of China under Grant No. 11722540,
the Fundamental Research Funds for the Central Universities,
Grants-in-Aid for Scientific Research (No. JP17K05441 (C)),
and
Grants-in-Aid for Scientific Research on Innovative Areas (No. 18H05407).

\appendix

%
\section{Mass spectrum of $S$-wave bottom baryons}\label{sec:sbottom}
%

Masses and decay constants of $S$-wave bottom baryons have been systematically investigated in Ref.~\cite{Liu:2007fg} using the method of QCD sum rules within HQET. In this paper we reevaluate their parameters. The results are summarized in Table.~\ref{tab:swave}, some of which are used to calculate decay widths of $P$-wave bottom baryons in Sec.~\ref{sec:sdecay} and ~\ref{sec:ddecay}. We refer to Refs.~\cite{Liu:2007fg,Chen:2017sci} for detailed discussions of these calculations.

For completeness, we also list their masses used in the present study, which are taken from PDG~\cite{pdg} and have been averaged over isospin:
\begin{eqnarray}
   \nonumber        \Lambda_{b}(1/2^+)  ~:~ m&=&5619.60 \mbox{ MeV} \, ,
\\ \nonumber            \Xi_{b}(1/2^+)  ~:~ m&=&5793.20 \mbox{ MeV} \, ,
\\ \nonumber       \Sigma_{b}(1/2^+)    ~:~ m&=&5813.4 \mbox{ MeV} \, ,
\\ \nonumber   \Xi_{b}^{\prime}(1/2^+)  ~:~ m&=&5935.02 \mbox{ MeV} \, ,
\\                     \Omega_b(1/2^+)  ~:~ m&=&6046.1 \mbox{ MeV}\, ,
\\ \nonumber   \Sigma_{b}^{*}(3/2^+)    ~:~ m&=&5833.6 \mbox{ MeV} \, , \, \Gamma=9.5 \mbox{ MeV} \, , \, g_{\Sigma_{b}^{*}\Lambda_b\pi}=5.85  \mbox{ GeV}^{-1} \, ,
\\ \nonumber        \Xi_{b}^{*}(3/2^+)  ~:~ m&=&5952.6 \mbox{ MeV} \, , \, \Gamma=1.28 \mbox{ MeV} \, , \, g_{\Xi_{b}^{*}\Xi_b\pi}=3.66  \mbox{ GeV}^{-1} \, ,
\\                   \Omega_b^*(3/2^+)  ~:~ m&=&6063 \mbox{ MeV} \, .
\end{eqnarray}

\begin{table}[hbt]
\begin{center}
\renewcommand{\arraystretch}{1.5}
\caption{Parameters of $S$-wave bottom baryons, extracted at $T=0.5 \mbox{ GeV}$. See Refs.~\cite{Liu:2007fg,Chen:2017sci} for detailed discussions. The flavor $\mathbf{\bar 3}_F$ bottom baryons of $J^P = 1/2^+$ composes one baryon multiplet where the spin of the two light quarks is $s_l=0$, and the flavor $\mathbf{6}_F$ bottom baryons of $J^P = 1/2^+$ and $3/2^+$ compose another baryon multiplet with $s_l=1$. See Refs.~\cite{Liu:2007fg,Chen:2017sci} for detailed discussions.}
\begin{tabular}{c | c | c | c | c | c}
\hline\hline
Multiplets & ~~~Baryon~~~ & ~~~Mass (MeV)~~~ & ~~~$\omega_c$ (GeV)~~~ & ~~~$\overline{\Lambda}$ (GeV)~~~ & ~~~$f$ (GeV$^{3}$)~~~
\\ \hline\hline
\multirow{3}{*}{$[\mathbf{\bar 3}_F,{1\over2}^+]$} & $\Lambda_b^0({1/2}^+)$ & $5637^{+68}_{-56}$ & $1.10$ & $0.77^{+0.14}_{-0.12}$ &
                                                     ~~~~~~~~$0.015^{+0.003}_{-0.002}$
\\
                                                   & $\Xi_b^0({1/2}^+)$     & $5780^{+73}_{-68}$  & $1.25$ & $0.91^{+0.12}_{-0.11}$ & ~~~~~~~~$0.020^{+0.004}_{-0.003}$
\\
                                                   & $\Xi_b^-({1/2}^+)$     & $5780^{+73}_{-68}$ & $1.25$ & $0.91^{+0.12}_{-0.11}$ & ~~~~~~~~$0.020^{+0.004}_{-0.003}$
\\ \hline
\multirow{6}{*}{$[\mathbf{6}_F,{1\over2}^+]$}    & $\Sigma_b^{+}({1/2}^+)$    & $5809^{+82}_{-76}$ & $1.30$ & $0.95^{+0.13}_{-0.12}$ & $\sqrt2 \times
                                                 \left(0.036^{+0.008}_{-0.006}\right)$
\\
                                                 & $\Sigma_b^{0}({1/2}^+)$    & $5809^{+82}_{-76}$  & $1.30$ & $0.95^{+0.13}_{-0.12}$ & ~~~~~~~~$0.036^{+0.008}_{-0.006}$
\\
                                                 & $\Sigma_b^{-}({1/2}^+)$    & $5809^{+82}_{-76}$ & $1.30$ & $0.95^{+0.13}_{-0.12}$ & $\sqrt2 \times \left(0.036^{+0.008}_{-0.006}\right)$
\\
                                                 & $\Xi_b^{\prime0}({1/2}^+)$ & $5903^{+81}_{-79}$  & $1.40$ & $1.04^{+0.11}_{-0.10}$ & ~~~~~~~~$0.044^{+0.009}_{-0.008}$
\\
                                                 & $\Xi_b^{\prime-}({1/2}^+)$ & $5903^{+81}_{-79}$  & $1.40$ & $1.04^{+0.11}_{-0.10}$ & ~~~~~~~~$0.044^{+0.009}_{-0.008}$
\\
                                                 & $\Omega_b^{-}({1/2}^+)$    & $6036^{+81}_{-81}$  & $1.55$ & $1.17^{+0.09}_{-0.09}$ & $\sqrt2 \times \left(0.057^{+0.011}_{-0.009}\right)$
\\ \hline
\multirow{6}{*}{$[\mathbf{6}_F,{3\over2}^+]$}    & $\Sigma_b^{*+}({3/2}^+)$ & $5835^{+82}_{-77}$  & $1.30$ & $0.95^{+0.13}_{-0.12}$ &$\sqrt{2\over3} \times
                                                  \left(0.036^{+0.008}_{-0.006}\right)$
\\
                                                 & $\Sigma_b^{*0}({3/2}^+)$  & $5835^{+82}_{-77}$  & $1.30$ & $0.95^{+0.13}_{-0.12}$ & $\sqrt{1\over3} \times \left(0.036^{+0.008}_{-0.006}\right)$
\\
                                                 & $\Sigma_b^{*-}({3/2}^+)$  & $5835^{+82}_{-77}$  & $1.30$ & $0.95^{+0.13}_{-0.12}$ & $\sqrt{2\over3} \times \left(0.036^{+0.008}_{-0.006}\right)$
\\
                                                 & $\Xi_b^{*0}({3/2}^+)$     & $5929^{+83}_{-79}$  & $1.40$ & $1.04^{+0.11}_{-0.10}$ & $\sqrt{1\over3} \times \left(0.044^{+0.009}_{-0.008}\right)$
\\
                                                 & $\Xi_b^{*-}({3/2}^+)$     & $5929^{+83}_{-79}$  & $1.40$ & $1.04^{+0.11}_{-0.10}$ & $\sqrt{1\over3} \times \left(0.044^{+0.009}_{-0.008}\right)$
\\
                                                 & $\Omega_b^{*-}({3/2}^+)$  & $6063^{+83}_{-82}$  & $1.55$ & $1.17^{+0.09}_{-0.09}$ & $\sqrt{2\over3} \times \left(0.057^{+0.011}_{-0.009}\right)$
\\ \hline \hline
\end{tabular}
\label{tab:swave}
\end{center}
\end{table}

\section{Sum rule equations}
\label{sec:othersumrule}

In this appendix we show the sum rule equations which are used to extract the $D$-wave decay properties of the $P$-wave bottom baryons belonging to the baryon doublet $[\mathbf{6}_F, 2, 1, \lambda]$.

The sum rule for $\Sigma_c^0[{3\over2}^-]$ belonging to $[\mathbf{6}_F, 2, 1, \lambda]$ is
\begin{eqnarray}
&& G_{\Sigma_c^0[{3\over2}^-] \rightarrow \Sigma_c^{*+}\pi^-} (\omega, \omega^\prime)
= { g_{\Sigma_c^0[{3\over2}^-] \rightarrow \Sigma_c^{*+}\pi^-} f_{\Sigma_c^0[{3\over2}^-]} f_{\Sigma_c^{*+}} \over (\bar \Lambda_{\Sigma_c^0[{3\over2}^-]} - \omega^\prime) (\bar \Lambda_{\Sigma_c^{*+}} - \omega)}
\\ \nonumber &=& \int_0^\infty dt \int_0^1 du \int \mathcal{D}\underline{\alpha} e^{i \omega^\prime t (\alpha_2 + u\alpha_3)} e^{i \omega t(1-\alpha_2-u\alpha_3)} \times 8 \times \Big (
\\ \nonumber &&
\frac{f_\pi v\cdot q}{24 \pi^2 t^2} \Phi_{4;\pi}(\underline{\alpha})
+ \frac{f_\pi v\cdot q}{24 \pi^2 t^2} \Psi_{4;\pi}(\underline{\alpha})
- \frac{ f_\pi v\cdot q}{24 \pi^2 t^2} \widetilde \Phi_{4;\pi}(\underline{\alpha})
- \frac{f_\pi v\cdot q}{24 \pi^2 t^2} \widetilde \Psi_{4;\pi}(\underline{\alpha})
- \frac{f_\pi u v\cdot q}{24 \pi^2 t^2} \Psi_{4;\pi}(\underline{\alpha})
+ \frac{f_\pi u v\cdot q}{24 \pi^2 t^2} \widetilde \Psi_{4;\pi}(\underline{\alpha}) \Big ) \, .
\end{eqnarray}

The sum rule for $\Xi_c^0[{3\over2}^-]$ belonging to $[\mathbf{6}_F, 2, 1, \lambda]$ is
\begin{eqnarray}
&& G_{\Xi_c^{\prime0}[{3\over2}^-] \rightarrow \Xi_c^{*+}\pi^-} (\omega, \omega^\prime)
= { g_{\Xi_c^{\prime0}[{3\over2}^-] \rightarrow \Xi_c^{*+}\pi^-} f_{\Xi_c^{\prime0}[{3\over2}^-]} f_{\Xi_c^{*+}} \over (\bar \Lambda_{\Xi_c^{\prime0}[{3\over2}^-]} - \omega^\prime) (\bar \Lambda_{\Xi_c^{*+}} - \omega)}
\\ \nonumber &=& \int_0^\infty dt \int_0^1 du \int \mathcal{D}\underline{\alpha} e^{i \omega^\prime t (\alpha_2 + u\alpha_3)} e^{i \omega t(1-\alpha_2-u\alpha_3)} \times 4 \times \Big (
\\ \nonumber &&
\frac{f_\pi v\cdot q}{24 \pi^2 t^2} \Phi_{4;\pi}(\underline{\alpha})
+ \frac{f_\pi v\cdot q}{24 \pi^2 t^2} \Psi_{4;\pi}(\underline{\alpha})
- \frac{ f_\pi v\cdot q}{24 \pi^2 t^2} \widetilde \Phi_{4;\pi}(\underline{\alpha})
- \frac{f_\pi v\cdot q}{24 \pi^2 t^2} \widetilde \Psi_{4;\pi}(\underline{\alpha})
- \frac{f_\pi u v\cdot q}{24 \pi^2 t^2} \Psi_{4;\pi}(\underline{\alpha})
+ \frac{f_\pi u v\cdot q}{24 \pi^2 t^2} \widetilde \Psi_{4;\pi}(\underline{\alpha}) \Big ) \, ,
\\ && G_{\Xi_c^{\prime0}[{3\over2}^-] \rightarrow \Sigma_c^{*+}K^-} (\omega, \omega^\prime)
= { g_{\Xi_c^{\prime0}[{3\over2}^-] \rightarrow \Sigma_c^{*+}K^-} f_{\Xi_c^{\prime0}[{3\over2}^-]} f_{\Sigma_c^{*+}} \over (\bar \Lambda_{\Xi_c^{\prime0}[{3\over2}^-]} - \omega^\prime) (\bar \Lambda_{\Sigma_c^{*+}} - \omega)}
\\ \nonumber &=& \int_0^\infty dt \int_0^1 du \int \mathcal{D}\underline{\alpha} e^{i \omega^\prime t (\alpha_2 + u\alpha_3)} e^{i \omega t(1-\alpha_2-u\alpha_3)} \times 4 \times \Big (
\\ \nonumber &&
\frac{f_K v\cdot q}{24 \pi^2 t^2} \Phi_{4;K}(\underline{\alpha})
+ \frac{f_K v\cdot q}{24 \pi^2 t^2} \Psi_{4;K}(\underline{\alpha})
- \frac{ f_K v\cdot q}{24 \pi^2 t^2} \widetilde \Phi_{4;K}(\underline{\alpha})
- \frac{f_K v\cdot q}{24 \pi^2 t^2} \widetilde \Psi_{4;K}(\underline{\alpha})
\\ \nonumber &&
- \frac{f_K u v\cdot q}{24 \pi^2 t^2} \Psi_{4;K}(\underline{\alpha})
+ \frac{f_K u v\cdot q}{24 \pi^2 t^2} \widetilde \Psi_{4;K}(\underline{\alpha}) \Big ) \, .
 \end{eqnarray}

The sum rule for $\Omega_c^0[{3\over2}^-]$ belonging to $[\mathbf{6}_F, 2, 1, \lambda]$ is
\begin{eqnarray}
&& G_{\Omega_c^0[{3\over2}^-] \rightarrow \Xi_c^{*+}K^{-}} (\omega, \omega^\prime)
= { g_{\Omega_c^0[{3\over2}^-] \rightarrow \Xi_c^{*+}K^{-}} f_{\Omega_c^0[{3\over2}^-]} f_{\Xi_c^{*+}} \over (\bar \Lambda_{\Omega_c^0[{3\over2}^-]} - \omega^\prime) (\bar \Lambda_{\Xi_c^{*+}} - \omega)}
\\ \nonumber &=&
\int_0^\infty dt \int_0^1 du e^{i (1-u) \omega^\prime t} e^{i u \omega t} \times 8 \times \Big (
\\ \nonumber &&
\frac{f_K v\cdot q}{24 \pi^2 t^2} \Phi_{4;K}(\underline{\alpha})
+ \frac{f_K v\cdot q}{24 \pi^2 t^2} \Psi_{4;K}(\underline{\alpha})
- \frac{ f_K v\cdot q}{24 \pi^2 t^2} \widetilde \Phi_{4;K}(\underline{\alpha})
- \frac{f_K v\cdot q}{24 \pi^2 t^2} \widetilde \Psi_{4;K}(\underline{\alpha})
\\ \nonumber &&
- \frac{f_K u v\cdot q}{24 \pi^2 t^2} \Psi_{4;K}(\underline{\alpha})
+ \frac{f_K u v\cdot q}{24 \pi^2 t^2} \widetilde \Psi_{4;K}(\underline{\alpha}) \Big ) \, .
\end{eqnarray}

The sum rule for $\Sigma_c^0[{5\over2}^-]$ belonging to $[\mathbf{6}_F, 2, 1, \lambda]$ is
\begin{eqnarray}
G_{\Sigma_b^-[{5\over2}^-] \rightarrow \Lambda_b^{0}\pi^-} (\omega, \omega^\prime)
= { g_{\Sigma_c^0[{3\over2}^-] \rightarrow \Sigma_c^{*+}\pi^-} f_{\Sigma_c^0[{3\over2}^-]} f_{\Sigma_c^{*+}} \over (\bar \Lambda_{\Sigma_c^0[{3\over2}^-]} - \omega^\prime) (\bar \Lambda_{\Sigma_c^{*+}} - \omega)}
= 0 \, .
\end{eqnarray}

%

%

\end{document}